# Impacts Towards a comprehensive assessment of the book impact by integrating multiple evaluation sources

Qingqing Zhou[1], Chengzhi Zhang[2, *]

*1. Department of Network and New Media, Nanjing Normal University, Nanjing 210023, China*
*2. Department of Information Management, Nanjing University of Science and Technology, Nanjing 210094, China*

**Abstract.**
The surge in the number of books published makes the manual evaluation methods (e.g. peer review) difficult to efficiently evaluate books. The use of books' citations and alternative evaluation metrics (e.g. library holdings, social media mentions, book reviews) can assist manual evaluation and reduce the cost of evaluation. However, most existing evaluation research was based on a single evaluation source with coarse-grained analysis, which may obtain incomprehensive or one-sided evaluation results of book impact. Meanwhile, relying on a single resource for book assessment may lead to the risk that the evaluation results cannot be obtained due to the lack of the evaluation data, especially for newly published books. Hence, this paper measured book impact based on an evaluation system constructed by integrating multiple evaluation sources. Specifically, we conducted finer-grained mining on the multiple evaluation sources, including books' internal evaluation resources (e.g. books' contents) and external evaluation resources (e.g. books' reviews, books' citations and books' usages). Various technologies (e.g. topic extraction, sentiment analysis, text classification) were used to extract corresponding evaluation metrics from the internal and external evaluation resources. Then, Expert evaluation combined with analytic hierarchy process was used to integrate the evaluation metrics and construct a book impact evaluation system. Finally, the reliability of the evaluation system was verified by comparing with the results of expert evaluation, detailed and diversified evaluation results were then obtained. The experimental results reveal that differential evaluation resources can measure the books' impacts from different dimensions, and the integration of multiple evaluation data can assess books more comprehensively. Meanwhile, the book impact evaluation system can provide personalized evaluation results according to the users' evaluation purposes. In addition, the disciplinary differences should be considered for assessing books' impacts.

**Keywords:** book impact assessment; multiple evaluation sources, review mining; citation context analysis; depth and breadth analysis

---

[*] Corresponding author: Chengzhi Zhang, E-mail: zhangcz@njust.edu.cn.

## 1 Introduction

With the rapid development of Internet and digitalization, people's reading and evaluation models of books are also changing. Literature databases, social media and e-commerce websites provide many new evaluation sources for book impact evaluation (Azer, 2019; Torres-Salinas et al., 2014). Meanwhile, the progress of digital storage and technologies about natural language processing provide technical support for measuring book impact. Therefore, the impact evaluation of books is no longer limited to the traditional evaluation metrics, such as peer reviews or citation frequencies. Massive alternative evaluation sources can be analyzed to detect more evaluation metrics (e.g. purchase intentions, citation functions) and thus overcome shortcomings of traditional metrics, such as high cost or time consumption (Torres-Salinas et al., 2017b; Zuccalá & Leeuwen, 2014). Hereby, currently, multiple evaluation resources have been used to assess impacts of books, including book contents (Mooney & Roy, 2000), book reviews (Chevalier & Mayzlin, 2006), book citations (Gorraiz et al., 2014b), book usages (Calhoun, 2011) etc. These books related evaluation resources can reflect the impacts of books from different dimensions, and provide supplementary information for the evaluation research from the corresponding dimensions.

However, most existing research was based on a single evaluation resource. The shortcomings of such evaluation method are obvious, as the used evaluation resource may be absent for some books, especially newly published books. For example, for 2739 books analyzed in (Kousha & Thelwall, 2016), only 84% books have google citations, 29% books have amazon reviews, and 7% books have Mendeley bookmarks. For 15928 books assessed in (Kousha et al., 2017), only 73.8% books have google citations, 34.6% books have Wikipedia citations, and 14.1% books have Goodreads reviews. Meanwhile, totally different or even contradictory evaluation results may be obtained by choosing different evaluation resources. For example, *Sentiment Analysis and Opinion Mining* by *Bing Liu* has been cited more than 5000 times in Google scholar, while it has only been discussed about 10 times in Amazon. The scientific integration of evaluation resources can not only solve these problems, but also provide comprehensive evaluation results for users without prior evaluation knowledge or users without obvious evaluation dimension tendency, so as to help users quickly obtain the evaluation conclusions they need (Torres-Salinas et al., 2017a). Hence, finer-grained mining on the multiple evaluation resources and the integration of corresponding evaluation results are necessary. This paper synthesized the multi-source evaluation data and then integrated metrics extracted from these sources to construct a multi-level and multi-dimensional evaluation metric system for assessing books' comprehensive impacts. The experimental results indicate that the integration of multiple evaluation sources can detect detailed evaluation information and meet users' personalized evaluation demands.

## 2 Related works

Currently, various resources are used to evaluate books' impacts. In this section, we describe two types of evaluation resources, namely books' external resources and internal resources.

Many external evaluation resources of books are used to evaluate the impacts of books, such as book reviews, book citations and book usages. Book reviews reflect users' direct attitudes on books (Zhang et al., 2019). Scholars analyze books' quality and evaluate values of books for scientific research with academic reviews (Gorraiz et al., 2014a; Zuccalá et al., 2014). For example, Kousha and Thelwall (2015) and Zhou and Zhang (2020b) measured books' impacts based on academic reviews from *Choice* and confirmed the validity of academic reviews for book impact evaluation.

Social media and e-commerce users post online reviews to express opinions on books' prices, papers, appearances etc. (Kousha & Thelwall, 2016). Online reviews from Amazon (Zhou et al., 2016) and Goodreads (Kousha et al., 2017; Maity et al., 2018) have been widely analyzed to identify impacts of books in different languages.

Citations of books are commonly used to assess books' impacts (Butler et al., 2017), and multiple citation databases provide extensive citation data for impact evaluation. Scopus (Zuccalá & Cornacchia, 2016), Web of Science Core Collection (Gorraiz et al., 2014b; Tsay et al., 2016), Google Scholar (Thelwall & Abrizah, 2014) and Microsoft Academic (Kousha & Thelwall, 2018) are effective evaluation resources. Meanwhile, Chinese Social Science Citation Index (Su et al., 2014) and Chinese Book Citation Index (Ye, 2014) are designed and developed for evaluating impacts of Chinese books. Books' citation literatures can also be systematically used for indicators of books' impacts. Zhou and Zhang (2020a) conducted fine-grained analysis on books' citation literatures to assess books' wider impacts. Meanwhile, citation contexts about books in citation literatures reveal researchers' citation intentions and attitudes on books. Mccain and Salvucci (2006) mined 574 citation contexts about The Mythical Man-Month to evaluate the its impact. Zhou and Zhang (2019) analyzed 2288 citation contexts about 370 books and then assessed impacts of these books.

With the development of Web 2.0, many alternative evaluation resources are mined and used for measuring books' use impact. Library holdings (White & Zuccalá, 2018), library loans (Cabezas-Clavijo et al., 2013), publisher prestige (Donovan & Butler, 2007), syllabus mentions (Kousha & Thelwall, 2008) and social media mentions (Batooli et al., 2016; Oberst, 2017) were extracted and analyzed to measure books' impacts from different aspects.

The above evaluation resources and metrics extracted from such resources are mainly based on books' external information. However, shortcomings of these external information cannot be ignored, as some books may not be commented or cited, the lack of evaluation data may result in the failure of evaluation. Hence, book impact assessment based on books' internal information is necessary. As the internal information of a book, the analysis of the book content, especially the full-text content, can reflect the quality of the book directly. However, due to the difficulty of obtaining books' contents, the evaluation analysis of books based on full texts is rare. Books' tables of contents are summaries of books' contents, researchers then used the tables of contents to measure the books' impacts in terms of the content dimension (Poulsen, 1996; Zhang & Zhou, 2020).

In conclusion, massive metrics extracted from various sources are proved to be useful for book impact assessment. The extracted metrics include both frequency-level metrics (e.g. citation frequencies and library holdings) and content-level metrics (e.g. metrics from reviews, citation contexts or tables of contents). Frequency-level metrics can provide intuitive evaluation results, while shortcomings of such metrics are obvious. Researchers cannot detect users' real reactions to books (e.g. whether users will recommend or buy books) or identify the applicable populations of books. Content-level metrics can overcome shortcomings of frequency-level metrics and reflect different impact dimensions from frequency information. In other words, metrics delivered from different sources cannot replace each other, but may play a complementary role. Integrating the existing evaluation resources reasonably and effectively to obtain books' comprehensive impacts is of great significance. Hence, this paper aims to integrate multi-source evaluation data to construct an evaluation system, so as to provide more detailed and comprehensive information for meeting the evaluation needs of different categories of users.

## 3 Research questions

Little research thus far has assessed book impacts based on a multi-source evaluation system constructed by integrating multiple resources, which may ignore book impacts in some dimensions, and then lead to the decline in the accuracy and practicability of evaluation results. Hence, the present study fills the gap by addressing the following research questions:

RQ1. Which metrics can reflect book impact more?

RQ2. Can the impacts of books be evaluated better by integrating multiple evaluation resources?

RQ3. Are there disciplinary differences in the book impact assessment?

## 4 Methodology

### *4.1 Framework*

The primary purpose of this paper is assessing books' comprehensive impacts by integrating multiple evaluation resources. We collect book evaluation resources from the internal and external dimensions of books. The internal evaluation resource is book content-related information, while the external evaluation resources of books include book review-, citation- and usage-related information. By mining and analyzing these resources (e.g. sentiment analysis, topic analysis), we can extract evaluation metrics of book impact and construct a book impact evaluation system. Then, we calculate weights and scores of each metric in the evaluation system, so as to get the impact results of books. In addition, we compare our evaluation results and scores evaluated by experts to verify the reliability of the assessment system. The overall framework is summarized in Figure 1.

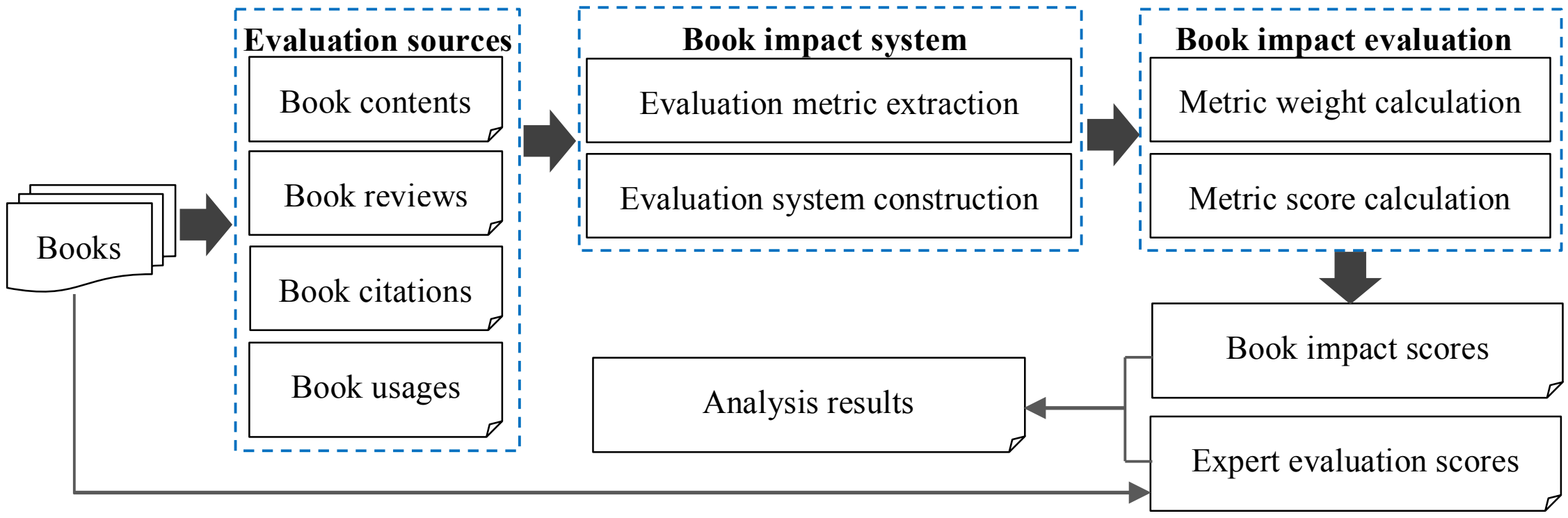

**Figure 1.** Framework of book impact assessment based multiple sources

### *4.2 Evaluation source collection*

This paper collects multiple evaluation resources to evaluate book impact from the internal and external dimensions of books, including book contents, reviews, citation information and usage information. These resources can directly reflect the attitudes and opinions to books of users related to book impacts (or users who pay attention to book impact evaluation), such as the authors, public readers, scholars and related institutions.

With the rapid development of e-commerce, people are more used to buy books online and generate massive book reviews. These reviews express users' opinions on books and reveal their sentiment tendencies on various aspects of books. The effective mining of reviews can identify users' purchase intentions and preferences. Meanwhile, online reviews are popular, massive, measurable and easy to access, which can be used as an important resource to evaluate impact of books (Zhou et al., 2016). Hence, for book reviews, we firstly matched Chinese discipline category (Standardization Administration of China, 2009) with book category provided by Amazon[1] to identify book disciplines (as the evaluation objects in this paper are Chinese books). Five disciplines were identified, including Computer Science, Literature, Law, Medicine and Sport Science. Then, we collected amazon reviews of books in the five disciplines in July 2017, and got 642258 reviews of 57627 books.

Books' tables of contents are summary of the books by authors, which abstract contents of books. Users can make a preliminary judgment on the contents of books by browsing the tables of contents (TOCs for short). Therefore, books' TOCs can be used to reflect impacts of books in contents. Hence, TOCs of the 57627 books were collected from amazon simultaneously for extracting content-related metrics.

Books' citation-related information includes books' citation frequencies and citation literatures (literatures that cited books). We extracted books' citation frequencies and citation literatures from Baidu Scholar[2] (one of the largest academic platform in the world with more than 1.2 billion academic resources[3]) with a crawler by matching titles, authors and publication years of books in August 2017. Then, citation frequencies and citation literatures (including titles, publication years, full texts) of 9757 books were collected (55467 of 65224 books had no citation). Meanwhile, we extracted citation contexts in citation literatures of books manually. Due to the high cost of manual annotation, we selected 500 books from the 9757 books according to the ratios of different citation frequencies. As part of citation literatures have no citation mark in the texts. Thus, we got 2288 citation contexts of 370 books. Each citation context contains five sentences, namely citation content and the former and latter two sentences of the citation content.

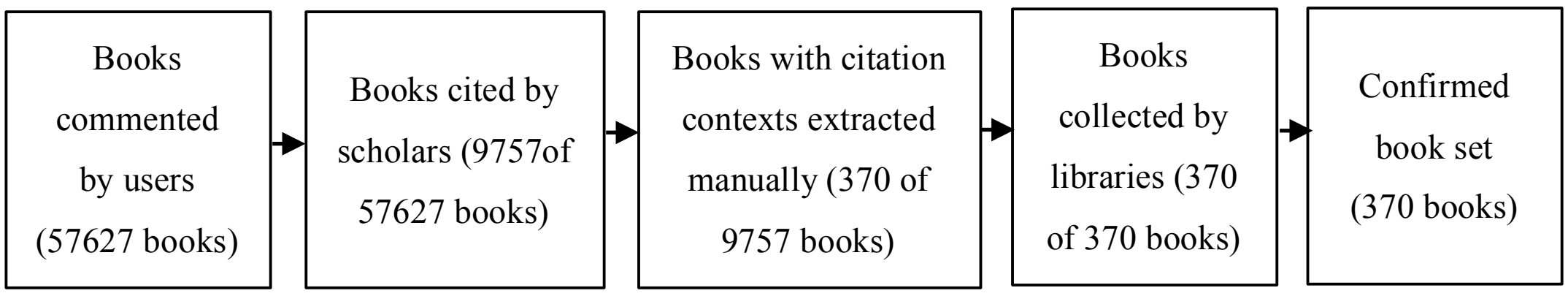

**Figure 2.** The process of data collection

**Table 1** Data statistics of books in five disciplines

| **Disciplines** | ***Computer Science*** | ***Literature*** | ***Law*** | ***Medicine*** | ***Sport Science*** | ***Total*** |
|---|---|---|---|---|---|---|
| #TOCs | 63 | 76 | 80 | 90 | 61 | 370 |
| #reviews | 2742 | 2891 | 1530 | 1879 | 1652 | 10694 |
| # citations | 385 | 404 | 450 | 506 | 332 | 2077 |
| # citation contexts | 284 | 548 | 614 | 585 | 257 | 2288 |
| #library holdings | 234 | 237 | 201 | 371 | 202 | 1245 |

Book usage information includes books' sales and library holdings. Due to Amazon's privacy rights, we cannot obtain the specific sale numbers of books in bulk. In this paper, we extracted book sale information from Amazon by matching ISBN of books, as Amazon provides books' sale ranking information on the book detail pages. We collected book' library holding information from

[1] https://www.amazon.cn/gp/book/all_category

[2] http://xueshu.baidu.com/

[3] https://xueshu.baidu.com/usercenter/show/baiducas?cmd=intro

WorldCat.org (OCLC). Finally, we obtained multi-dimensional evaluation information of 370 Chinese books (published from 1985 to 2016). The process of data collection is shown in Figure 2. Data statistics are shown in Table 1.

### *4.3 Construction of evaluation metric system for book impact*

We constructed the evaluation system of book impact with four resources: book contents, book reviews, book citations and book usages. We firstly conducted data mining on the multiple evaluation resources, including multi-granularity sentiment analysis, depth and breadth analysis, and citation context analysis, so as to obtain corresponding evaluation metrics. Then, an impact evaluation system was obtained based on the demonstration by domain experts.

#### *4.3.1 Impact assessment metrics from book contents*

This paper analyzed books' TOCs to measure book impacts from the dimension of book contents. Specifically, we conducted topic analysis on books' TOCs with LDA (Latent Dirichlet Allocation) to calculate books' depth and breadth (Hoffman et al., 2010; Pons-Porrata et al., 2007). We held that books introduced less topics tend to be more insightful, while books with more uniformly topic distributions may get higher breadth scores (Zhang & Zhou, 2020).. Then, we got two evaluation metrics, including TOC depth and TOC breadth, as shown in Figure 3. TOC depth refers to the depth of book contents reflected in the books' TOCs, while TOC breadth refers to the breadth of book contents reflected in the books' TOCs. The two metrics can be computed by equation (1) and (2).

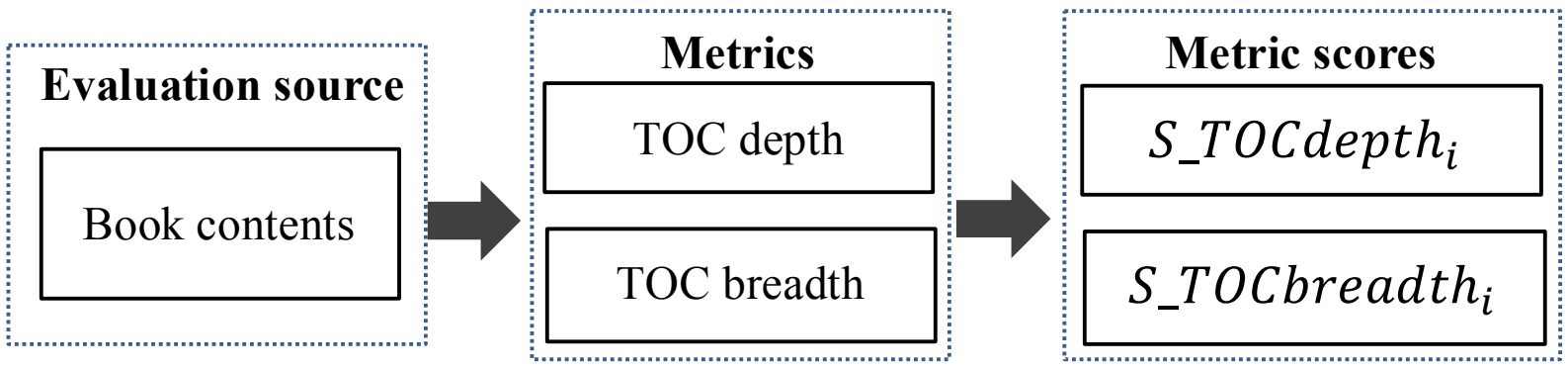

**Figure 3.** Impact assessment metrics from book contents

$$S_TOCdepth_i = 1/(\frac{\#TOCtopics_i}{\#pages_i}) \quad (1)$$

$$S_TOCbreadth_i = -\frac{1}{\ln(\#TOCtopics_i)} \sum_{j=1}^{\#TOCtopics_i} p_TOCtopics_{ij} \, ln(p_TOCtopics_{ij}) \quad (2)$$

Where, $S_TOCdepth_i$ means depth score of book $i$, $\#TOCtopics_i$ is number of topics expressed in the table of contents of book $i$, $\#pages_i$ means pages of the book $i$. $S_TOCbreadth_i$ denotes breadth score of book $i$, $p_TOCtopics_{ij}$ is the topic probability of the book $i$ in topic $j$.

#### *4.3.2 Impact assessment metrics from book reviews*

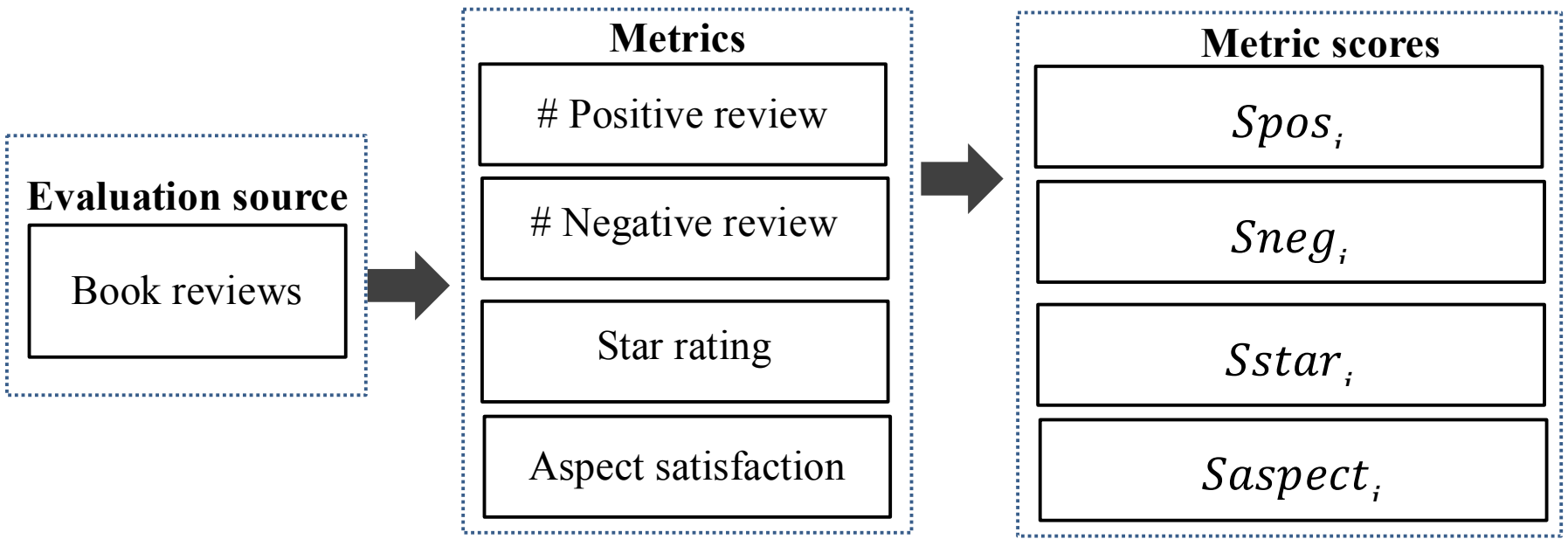

**Figure 4.** Impact assessment metrics from book reviews

Book reviews reflect users' opinions on books and books' aspects, such as *price*, *printing*, and *paper*. Hence, in order to get users' overall sentiments and aspect sentiments, we conducted multi-granularity sentiment analysis on book online reviews (Book reviews in this paper refer to online reviews of books. We did not analyze books' scholar reviews published in journals, as the number of books in the corpus commented by scholars is too small, accounting for only about 18.38%.)(Zhou et al., 2016). Specifically, we used supervised machine learning to identify the sentiment polarities of reviews. Then, we extracted aspects of books via deep learning (i.e. Word2Vec[4]) and detected sentiment polarities of aspects in each review (Zhou & Zhang, 2018). Hereby, four evaluation metrics were extracted from book reviews, including the number of positive reviews, number of negative reviews, star rating and aspect satisfaction, as shown in Figure 4. Aspect satisfaction reflects users' satisfactions on aspects of books. Scores of the four metrics can be compute with equation (3) to (7).

$$Spos_i = \#pos_i \tag{3}$$

Where, $Spos_i$ is the score of the positive review metric of book $i$; $\#pos_i$ is the number of positive reviews of book $i$.

$$Sneg_i = \#neg_i \tag{4}$$

Where, $Sneg_i$ is the score of the negative review metric of book $i$; $\#neg_i$ is the number of negative reviews of book $i$.

$$Sstar_i = \sum_{j=1}^{n_i} star_{ij} \Big/ n_i \tag{5}$$

Where, $Sstar_i$ denotes the star rating score of book $i$, $n_i$ means numbers of reviews of book $i$, $star_{ij}$ means the star rating in review $j$ of book $i$.

$$Saspect_i = \sum_{j=1}^{m_i} aspect_{ij} \Big/ m_i \tag{6}$$

$$aspect_{ij} = \sum_{k=1}^{n_{ij}} v_{ijk} \Big/ \sum_{k=1}^{n_{ij}} |v_{ij}| \tag{7}$$

Where, $Saspect_i$ denotes the aspect satisfaction score of book $i$, $aspect_{ij}$ means score of aspect $j$ about book $i$, $m_i$ means the number of aspects about book $i$. $v_{ijk}$ denotes aspect score of aspect $j$ in review $k$ about book $i$. If aspect $j$ in review $k$ is positive, $v_{ijk}$ equals 1, else it equals -1. $n_{ij}$ means the number of reviews with aspect $j$ about book $i$.

[4] http://word2vec.googlecode.com/svn/trunk/

*4.3.3 Impact assessment metrics from book citations*

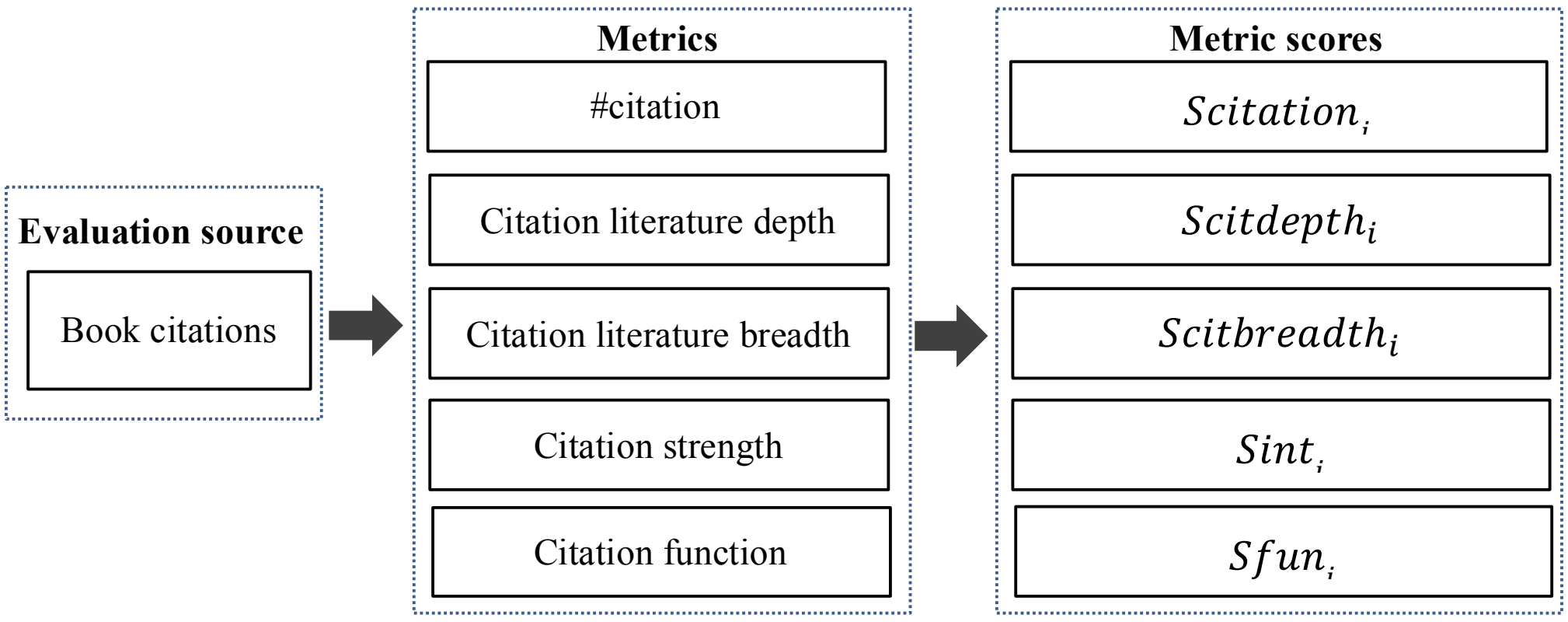

**Figure 5.** Impact assessment metrics from book citations

We extracted citation-based metrics from two citation sources, including citation frequency and citation literature. The citation frequency of books reflects scholars' opinions and attitudes on books. Generally, books with higher citation frequencies tend to get higher impacts (Kousha et al., 2011). For citation literatures, we can analyze the depth and breadth of books' citation literatures to measure books' depth and breadth (Zhou & Zhang, 2020a). Meanwhile, the analysis on citation contexts in citation literatures can identify citation intentions of scholars, which can measure detailed impacts of books (Zhou & Zhang, 2019). Hence, we can get five evaluation metrics from book citations, including citation frequency, citation literature depth, citation literature breadth, citation intensity and citation function, as shown in Figure 5. Citation literature depth means the depth of a book reflected by literatures cited the book, while citation literature breadth means the breadth of a book reflected by literatures cited the book. Citation function refers to scholars' purposes of citing books, including background citation, comparison citation and use citation (Hernández-Alvarez et al., 2017). Background citation means the book is cited to elaborate the frontier value, theoretical significance or practical value of a research field from a macro perspective. Comparison citation is cited for comparing the theories, methods, results or conclusions from books with the authors' research. Use citation aims to cite theories, methods, data, tools, etc. from existing books. Citation intensity denoted citation frequencies of a book in one citation literature.

For calculating scores of the five metrics, we conducted finer-grained analysis on the citation resources. Specifically, we counted numbers of citation literatures to get scores of citation frequencies, which can be calculated by equation (8).

$$Scitation_i = \#citation_i \tag{8}$$

Where, $Scitation_i$ is the score of the citation frequency metric of book $i$; $\#citation_i$ is the number of citations of book $i$.

We extracted topics expressed by citation literatures to reflect depth and breadth of books from the dimension of book citation. We held that books with more citation literatures and the citation literatures introduced fewer topics tend to get higher depth scores. Meanwhile, books with more uniformly topic distributions tend to get higher breadth scores. Hence, the depth and breadth of books based on citation literatures can be computed by equation (9) and (10).

$$Scitdepth_i = \frac{\#citation_i}{\#cittopics_i} \tag{9}$$

Where, $Scitdepth_i$ means the depth score of book $i$ based on citation literatures, $\#cittopics_i$ is topic numbers expressed in citation literatures of book $i$, $\#citation_i$ means citation frequency of book $i$, i.e. numbers of citation literatures of book $i$.

$$Scitbreadth_i = -\frac{1}{\ln(\#cittopics_i)}\sum_{j=1}^{\#cittopics_i} p_cittopics_{ij} ln\,(p_cittopics_{ij}) \quad (10)$$

Where, $Scitbreadth_i$ denotes the breadth score of book $i$ based on citation literatures, $\#cittopics_i$ is the number of topics of book $i$, $p_cite_topics_{ij}$ is the topic probability of the book $i$ in topic $j$.

We counted citations about a given book in a citation literature to calculate citation intensity of the book, which can be computed by equation (11)

$$Sint_i = \frac{\sum_{j=1}^{n} int_{ij}}{Scitation_i} \quad (11)$$

Where, $Sint_i$ denotes citation intensity score of book $i$, $int_{ij}$ means citation intensity score of book $i$ in citation literature $j$, $Scitation_i$ is citations of book $i$.

We conducted text classification on citation contexts extracted from citation literatures to identify scholars' three different citation functions, and then calculated metric scores of citation function with equations (12) and (13) (Hernández-Alvarez et al., 2017).

$$Sfun_i = \frac{\sum_{j=1}^{n} fun_{ij}}{n_i} \quad (12)$$

$$fun_{ij} = \begin{cases} 1, \text{ Background citation} \\ 2, \textit{Comparison citation} \\ 3, \textit{ Use citation} \end{cases} \quad (13)$$

Where, $Sfun_i$ denotes citation function score of book $i$, $fun_{ij}$ means citation function score of the $j$th citation context about book $i$. $n_i$ is the total citation frequency in the texts of citation literatures about book $i$.

*4.3.4 Impact assessment metrics from book usages*

The usages of books (e.g. library holdings and sales) are closely related to books' use impacts. Books with more library holdings and sales may get higher impacts (White et al., 2009). Therefore, in terms of book usages, we extracted four metrics, including library holding number, library holding region, library holding distribution and sale, as shown in Figure 6. Library holding numbers is the total number of a book in libraries around the world. Library holding region measures how many countries collect the book. Library holding distribution refers to holding distribution of the book in libraries. The four usage-related metrics can by equations (14) to (17).

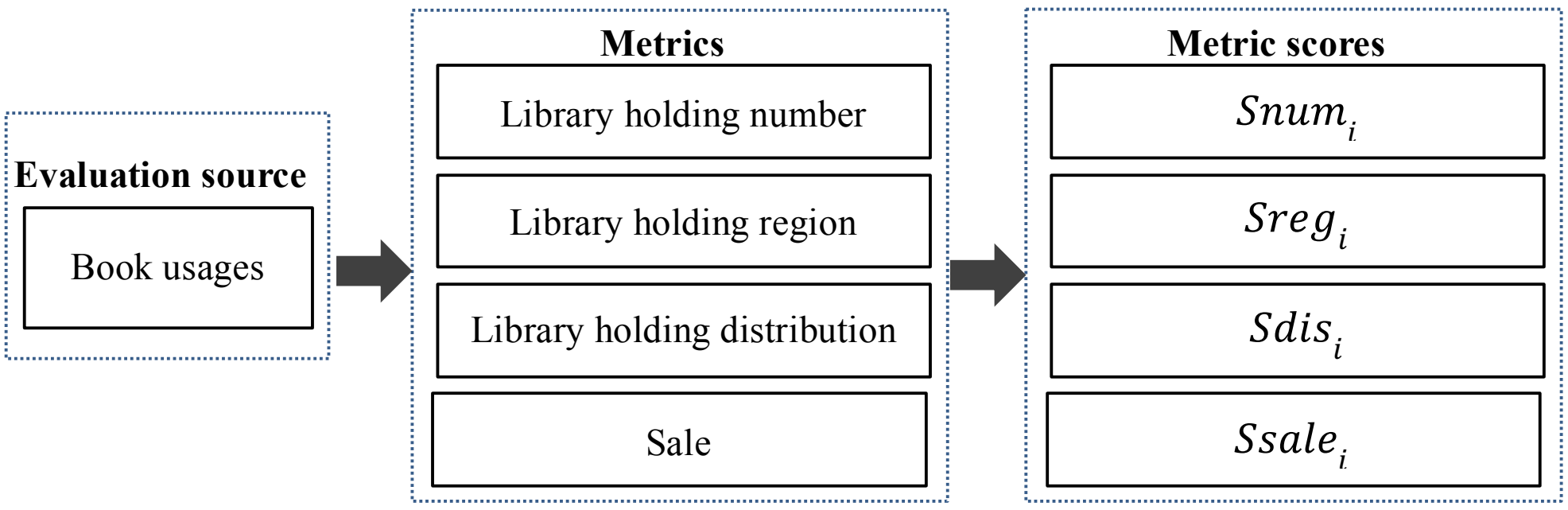

**Figure 6.** Impact assessment metrics from book usages

$$Sreg_i = \#holdreg_i \tag{14}$$

$$Snum_i = \sum_{j=1}^{Sreg_i} \#holdnum_{ij} \tag{15}$$

$$Sdis_i = -\frac{1}{\ln(Sreg_i)} \sum_{j=1}^{Sreg_i} p_holdings_{ij} \ln(p_holdings_{ij}) \tag{16}$$

$$Ssale_i = \#sale_i \tag{17}$$

Where, $Sreg_i$ is the score of holding regions of book $i$; $\#holdreg_i$ is the number of regions that collected book $i$. $Snum_i$ is the score of holding numbers of book $i$; $\#holdnum_{ij}$ is the number of library holdings of book $i$ in region $j$. $Sdis_i$ is the score of holding distributions of book $i$, $p_holdings_{ij}$ is the probability of the book $i$ in region $j$. $Ssale_i$ denotes the score of sale of book $i$; $\#sale_i$ is the reordered sales ranking of book $i$.

### *4.4 Calculation of metric weights for book impact assessment*

Based on the above analysis, we constructed a multi-level and multi-dimensional book impact evaluation system, as shown in Figure 7. Each metric can be quantified to reflect different characteristics of books and be used to evaluate the impact of books.

Expert evaluation combined with analytic hierarchy process (AHP) was used to calculate weights of evaluation metrics (Saaty, 2005). The AHP decomposes the problem into different factors according to the requirements of the overall goal. Based on the interrelated influence among factors, the factors are aggregated and combined at different levels to form a multi-level structure model. Finally, the problem comes down to the determination of the relatively important weights of the lowest level (i.e. evaluation metrics) relative to the highest level (i.e. book evaluation). Therefore, AHP is effective for hierarchical decision analysis, and can be used to calculate the weights of metrics in the evaluation system (Lee & Kozar, 2006).

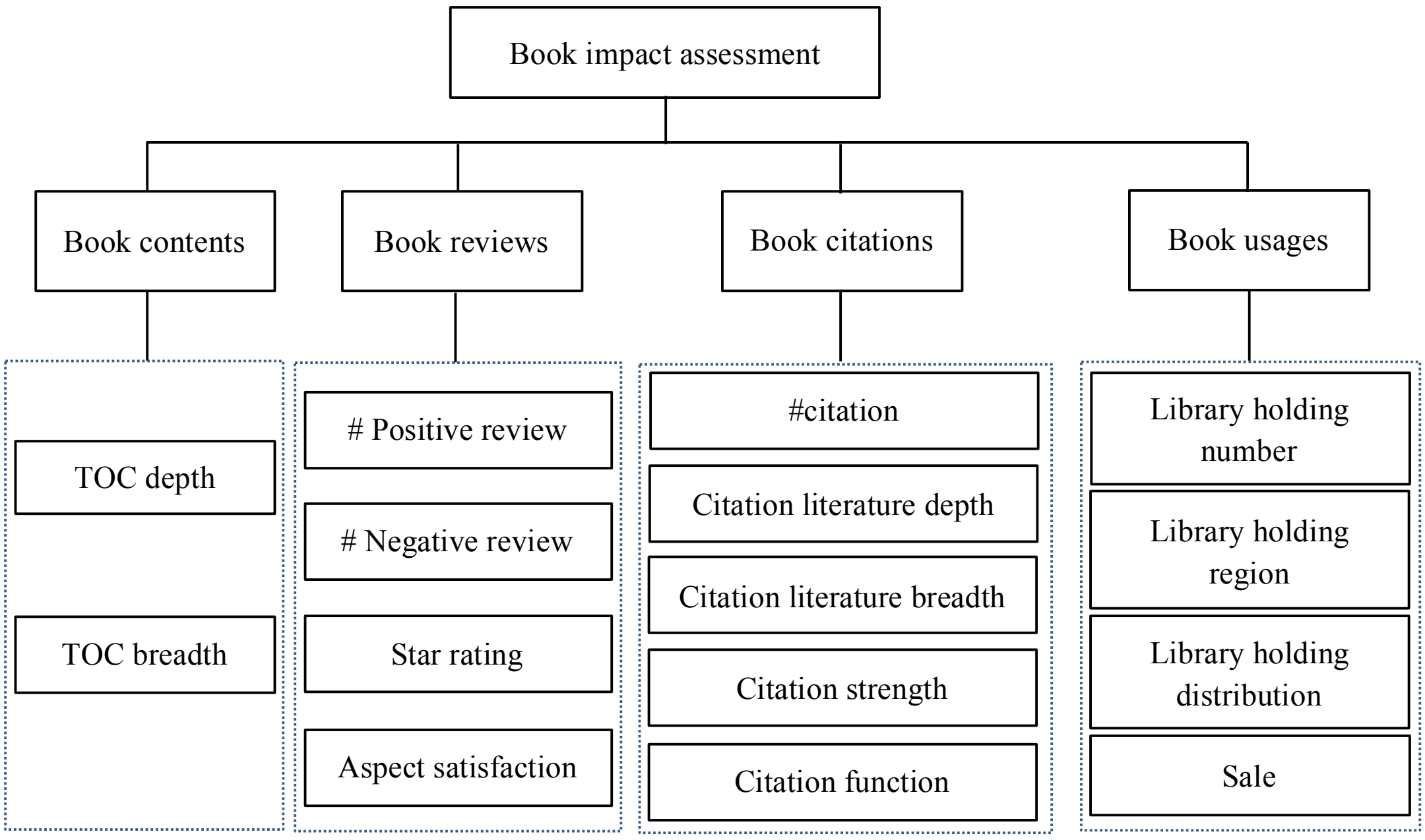

**Figure 7.** Book impact assessment system

Firstly, we invited experts in the field of book impact assessment (including scholars and relevant practitioners) to participate in the metric importance survey, so as to obtain the initial weights of metrics. 65 questionnaires were sent out and 53 valid questionnaires are collected. The questionnaire is shown in Appendix A. We use the 5-level scale to evaluate importance of metrics, ranging from 1 for "very unimportant" to 5 for "very important". Then, we get initial weights of all metrics in Figure 7. Finally, based on the results of the questionnaire survey, AHP was used to calculate the final weights of all metrics (Cheng & Li, 2001).

### *4.5 Calculation of book impact scores*

We integrated the evaluation metrics of multiple evaluation sources to determine the book impact score. Specifically, we normalized the score of each metric, and then book impact scores were obtained by weighted sum of the normalized scores with equation (18) and (19).

$$Score_i = \sum_{j=1}^{m}(NorS_{ij} * w_j) \tag{18}$$

$$NorS_{ij} = 2 * \text{atan}(S_{ij}) / \pi \tag{19}$$

Where, $w_j$ denotes weighting of metric $j$, $m$ is the number of metrics, $NorS_{ij}$ is normalized score of metric $j$ about book $i$. $S_{ij}$ is score of metric $j$ about book $i$.

## 5 Results

### *5.1 Analysis on metric weights of book impact assessment*

In order to determine which metric is more important for measuring book impacts (i.e. for answering RQ1), we calculated the weights of different metrics in the evaluation system. Figure 8 shows the weight scores of primary metrics. Figure 8 (a) presents the initial importance of the four primary metrics scored by 53 experts, and Figure 8 (b) reports the final weight scores of the four primary metrics. We can see from Figure 8 that the weight of book content is slightly higher than the other three metrics. It indicates that the importance of the four first-class metrics for book impact evaluation is close, while the book content is relatively more important. Meanwhile, the evaluation results from experts reveal that the first-class evaluation metrics extracted from four evaluation resources can be used to measure book impact. These metrics assess books' impacts of different dimensions from the internal and external aspects of books. Therefore, the integration of the four evaluation dimensions (or four evaluation resources) can be used to comprehensively evaluate the impacts of books.

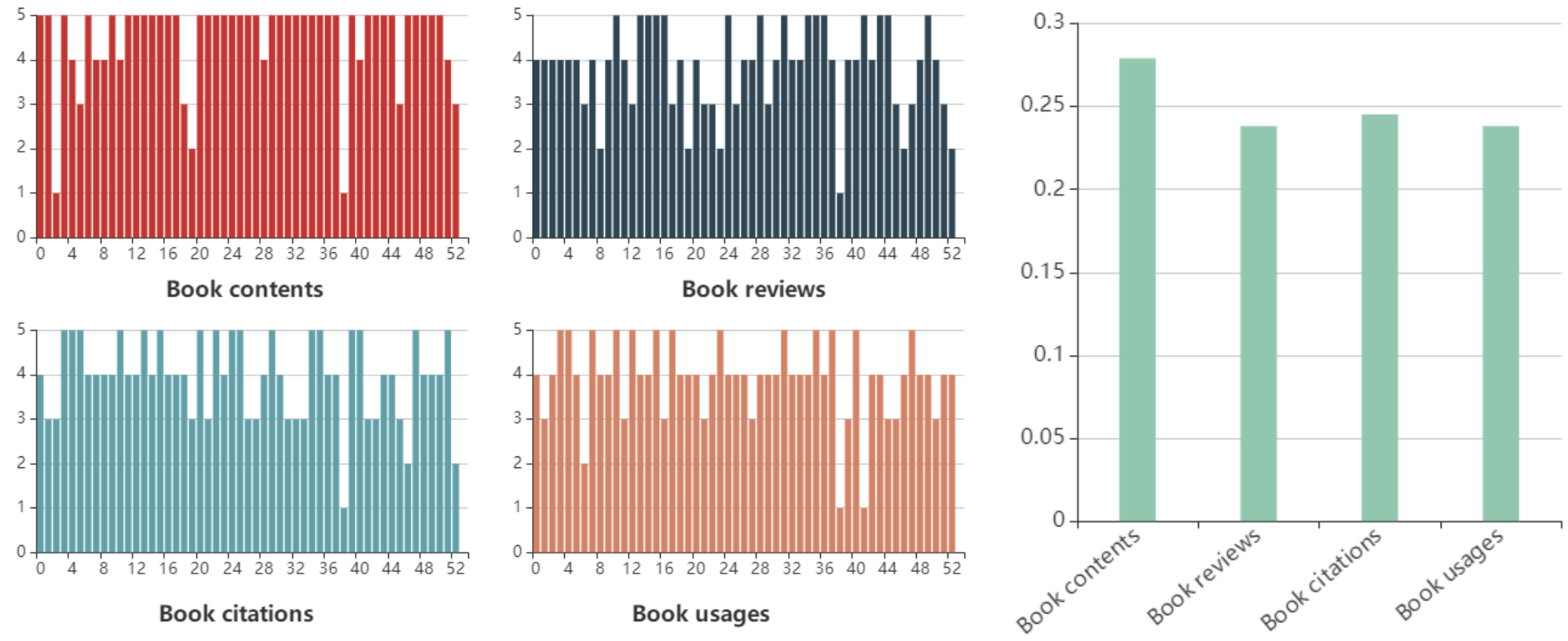

(a) (b)

**Figure 8.** The weight scores of primary metrics

Table 2 represents weights of secondary evaluation metrics in the book impact assessment system. For the secondary metrics, the weights of the internal evaluation metrics (i.e. the metrics extracted from the book content) are similar, about 0.14. The weights of the external evaluation metrics (i.e. the metrics extracted from book review, book citation and book usage) distribute between 0.047 and 0.064 and lower than the internal evaluation metrics. It reflects that book content is a quite important book evaluation resource. However, the existing research on book impact assessment is rarely based on book content. This may because books' contents often cannot be easily obtained online, and the difficulty of content analysis or processing is obviously higher than that of academic articles and other types of publications. In addition, the sum of the evaluation metrics weights from the outside of books (0.7211) is higher than internal evaluation metrics (0.2789). It indicates that the impact evaluation of books cannot only be based on the internal evaluation metrics, various external evaluation metrics are also an important evaluation basis. In summary, we can only obtain books' impacts from one dimension if we based on a single data source, and once there is a lack of data in this dimension (e.g., no book reviews), the impacts of books cannot be evaluated. Therefore, integrating multi-source data to evaluate the impacts of books can effectively avoid such shortcomings, and provide comprehensive evaluation results for users.

**Table 2.** The weights of book impact evaluation metrics

| Primary metrics | Secondary metrics | Weights of secondary metrics |
|---|---|---|
| Book contents | TOC depth | 0.1443 |
| | TOC breadth | 0.1346 |
| Book reviews | #positive review | 0.0640 |
| | #negative review | 0.0622 |
| | Star rating | 0.0578 |
| | Aspect satisfaction | 0.0540 |
| Book citations | #citation | 0.0502 |
| | Citation literature depth | 0.0498 |
| | Citation literature breadth | 0.0477 |
| | Citation strength | 0.0491 |
| | Citation function | 0.0482 |
| Book usages | Library holding number | 0.0598 |
| | Library holding region | 0.0569 |
| | Library holding distribution | 0.0578 |
| | Sale | 0.0636 |

Figure 9 shows the metric score ranks of 5 books with the highest impact scores. We can see score ranks of the 5 books in the 15 metrics are varied. It reveals that even books with high impacts are difficult to get high scores in all dimensions. Meanwhile, it also indicates that book impact evaluation based on a single evaluation resource may get one-sided evaluation results.

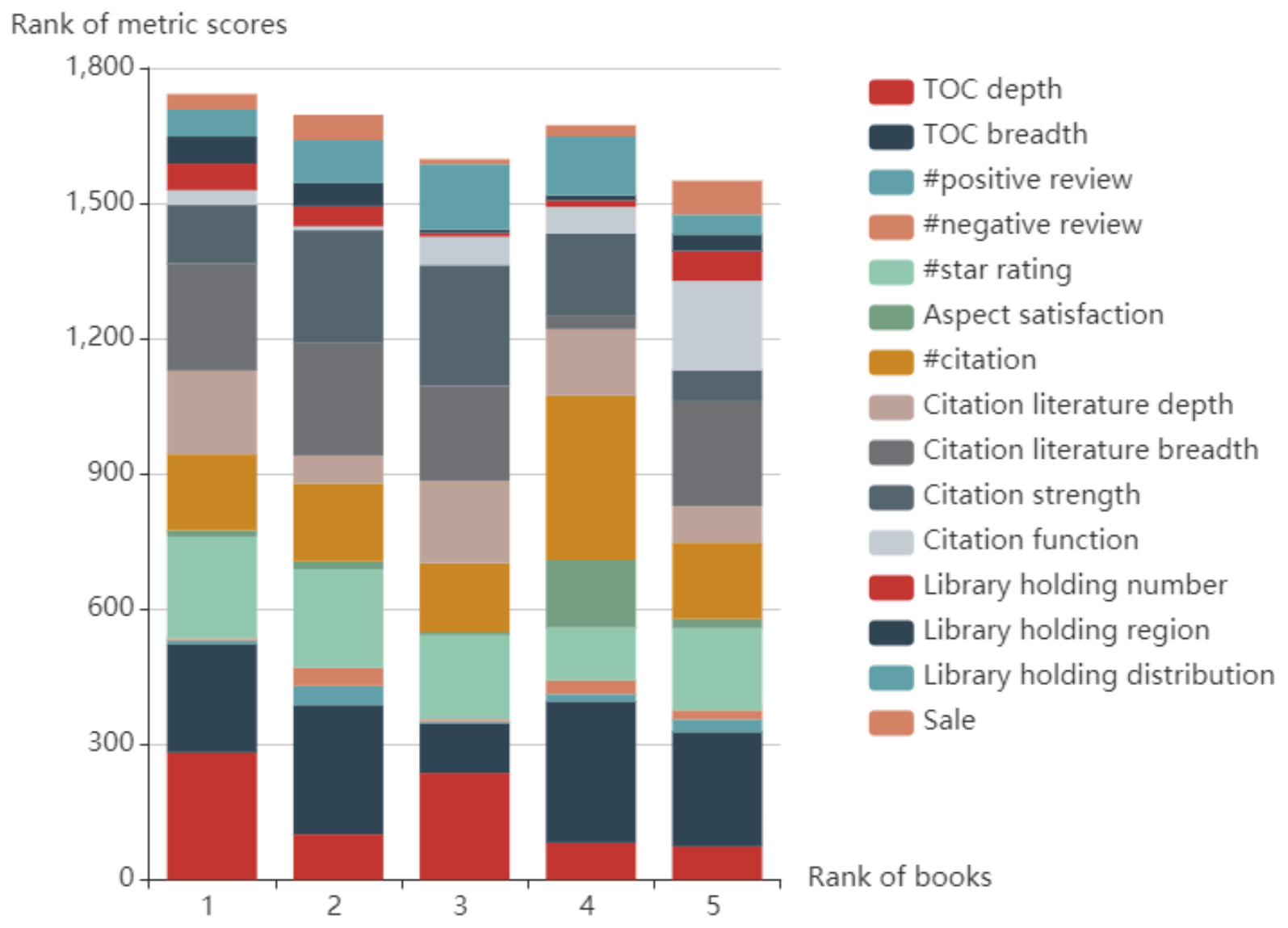

**Figure 9.** Metric score ranks of Top 5 books

### *5.2 Analysis on impact scores of book impact assessment*

#### *5.2.1 Reliability analysis on book impact assessment results*

In order to verify the reliability of the book impact results based on the impact evaluation system (i.e. for answering RQ2), we invited experts to evaluate the books' impacts manually, and then compared the two evaluation results. Specifically, we firstly took 48 books in 8 research domains of computer science and 30 books in 5 research domains of literature as experimental samples, as shown in Table 3. Then, we invited experts in the field of computer science and literature to manually assess the importance of books in corresponding disciplines by using a 5-level scale, ranging from 1 for "low impact" to 5 for "high impact". Meanwhile, we provided detailed page links of books on Amazon and Douban book[5] (an online book reading and comment website) for respondents to understand books. The questionnaire of books in literature is shown in Appendix B (The questionnaire of books in computer science is similar). 56 valid questionnaires related to computer science and 48 valid questionnaires related to literature were collected from experts. In the valid questionnaires, more than 80% of the respondents have master's degree or above, of which about 30% are doctors. Thirdly, we calculated the average score of expert evaluation as the final impact score of each book. Finally, we conducted correlation analysis between expert evaluation scores (i.e. book impact based on manual evaluation) and automatic assessment scores (i.e. book impact based on evaluation metric system). The results are shown in Table 4. It can be seen from Table 4 that the automatic book impact scores have a significant positive correlation with the expert evaluation results. It indicates that the calculation results based on our evaluation system are reliable.

[5] https://book.douban.com/

**Table 3.** Domains and numbers of books for expert evaluation

| Disciplines | Domains | #books | Domains | #books |
|---|---|---|---|---|
| Computer Science | Computer control simulation and artificial intelligence | 10 | Software engineering | 5 |
| | Computer network security | 7 | Programming and development | 7 |
| | Database | 5 | PLC Technology | 3 |
| | Operating system | 6 | Computer algorithms | 5 |
| Literature | Literature research | 7 | Prose | 5 |
| | Novel | 6 | History | 3 |
| | Poetry and Drama | 9 | | |

**Table 4.** Correlations between book comprehensive impact scores and expert evaluation scores

| Disciplines | Spearman correlation coefficients | N |
|---|---|---|
| Computer science | $0.631^{**}$ | 48 |
| Literature | $0.715^{**}$ | 30 |

Note: **. Significant at p=0.01

*5.2.2 Impact scores of book impact assessment*

Based on the multi-source data mining and analysis, we got the book impact assessment results, as shown in Figure 10. From Figure 10 we can see scores of books' comprehensive impacts range from 0.39 to 0.66, and most books are lower than 0.6. It indicates that the number of books with high impacts is relatively small, and most of them are in the set of low impact. Hence, books related scholars and institutions need to allocate resources effectively, as books cannot always get high scores in all aspects.

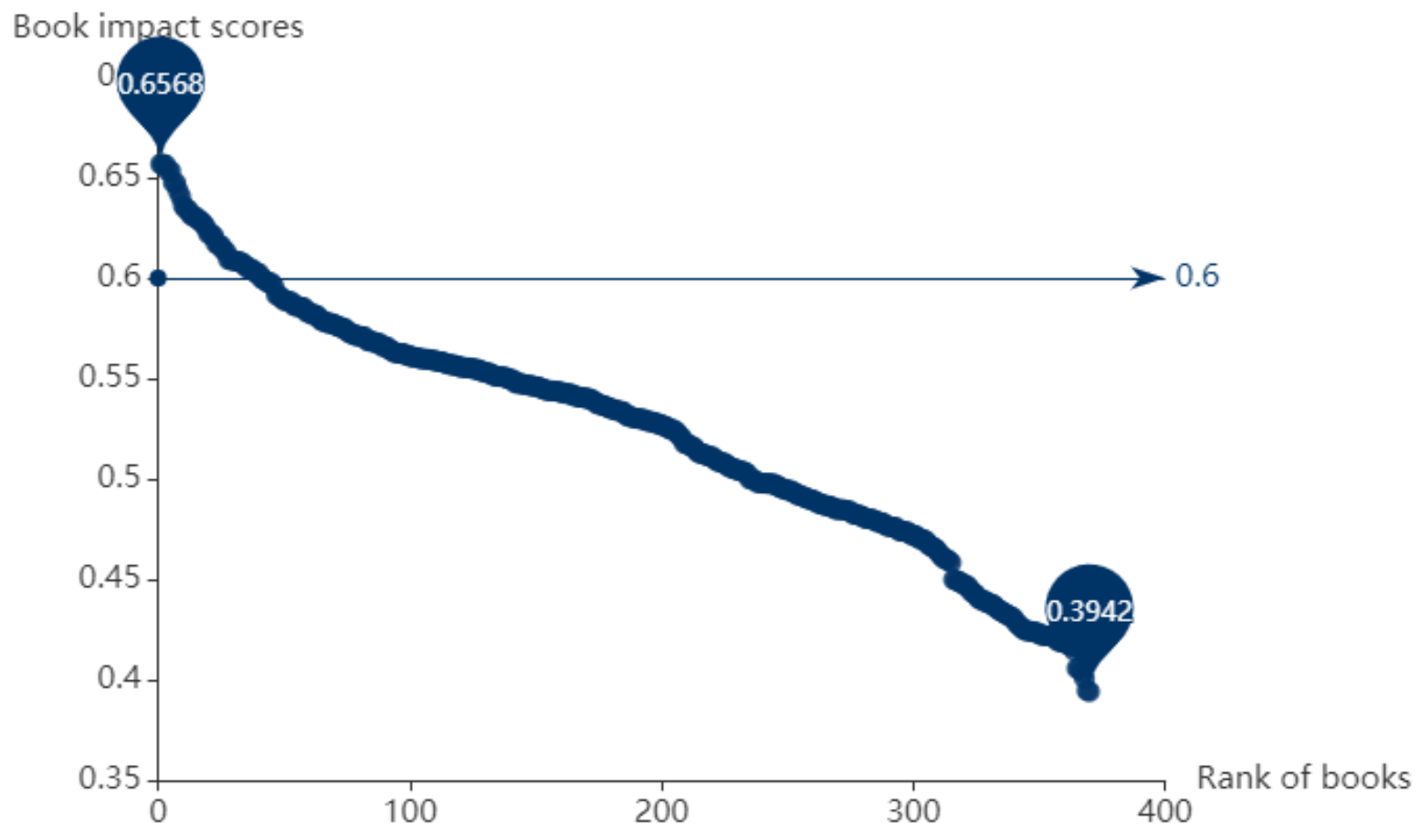

**Figure 10.** Scores of book impact assessment

*5.3 Discipline analysis on book impact assessment results*

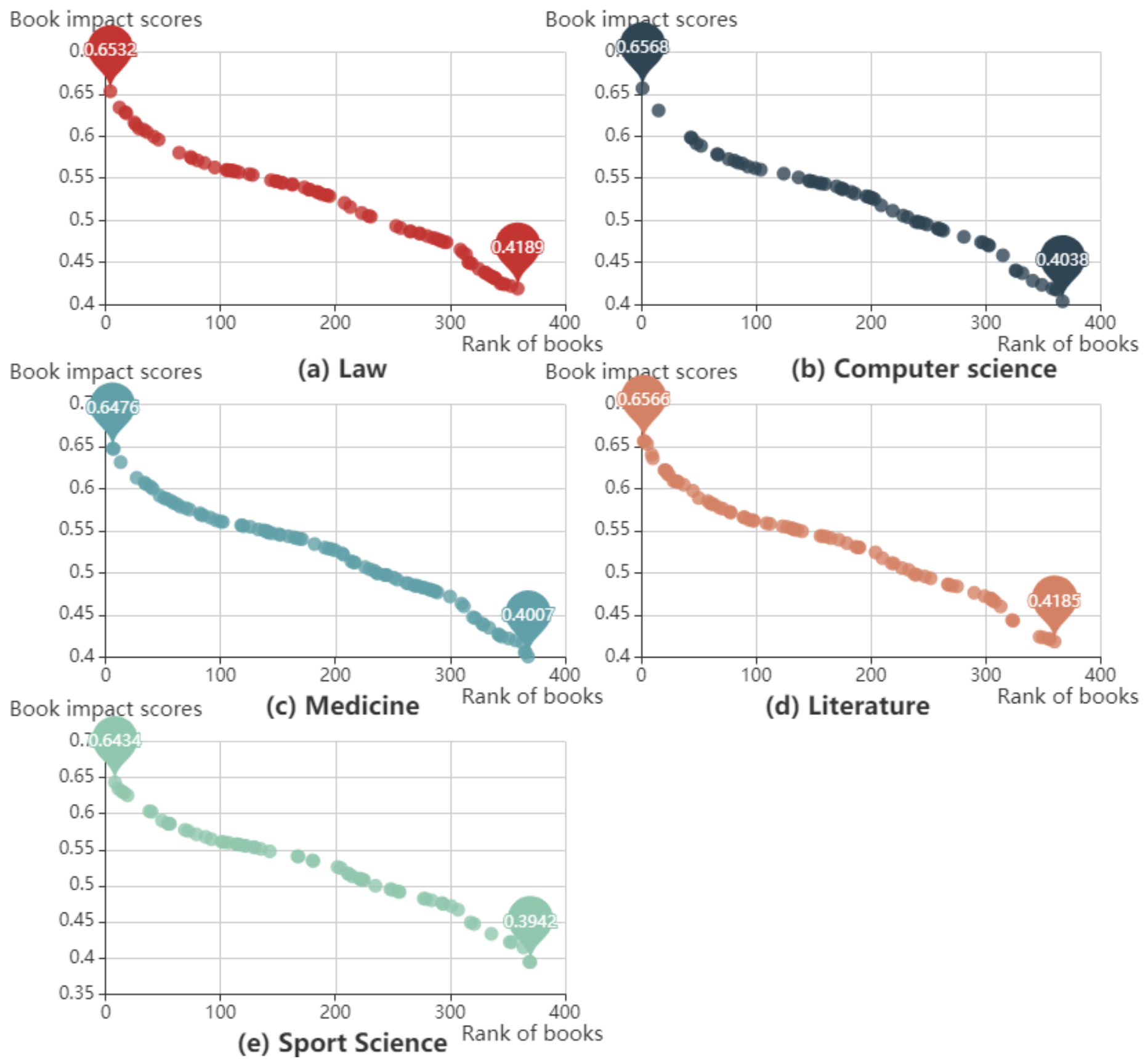

**Figure 11.** Scores of book impacts in different disciplines

In order to identify the disciplinary differences (i.e. for answering RQ3), we counted the book impacts scores in different disciplines and identified their score interval distributions. Figure 11 shows the impact scores of books in five disciplines. It can be seen from Figure 11 that the distribution trends of book impact scores in different disciplines are similar. There are less books in the high score area or low score area of each discipline, and most books are concentrated in the middle area. However, the impact scores of different disciplines are quite different. Law, computer science and literature get book impact scores higher than 0.65, while impact scores of books in medicine and sport science are all lower than 0.65. In addition, the number of books with impact scores higher than 0.6 in computer science is significantly less than that in other four disciplines, and only books in sport science get impact scores lower than 0.4. Hence, we can conclude that that disciplinary differences are existing, and users (including individual users and institutional users) need to consider the disciplinary differences when selecting, comparing and awarding books.

We counted the number distributions of different disciplines in different book impact score intervals, as shown in Figure 12. The impact scores of most books are in the middle score interval (i.e. 0.4-0.6). Meanwhile, about 10% books get impact scores higher than 0.6, while less than 1% books get impact scores lower than 0.4. The distribution results are consistent with the above analysis results based on Figure 10. In terms of discipline differences, we can see that the proportion of sports science books in low score interval (i.e. 0.3-0.4) is significantly higher than that of other disciplines. In the middle score interval, the proportions of books in law and medicine are higher. The proportion of literature in high score interval (i.e. 0.6-0.7) is highest, while the number of computer science books in high score interval is least. The proportion difference of the five disciplines in the four impact intervals indicates that there are obvious disciplinary differences in the distribution of the impact scores, especially the distributions of the extreme impact scores.

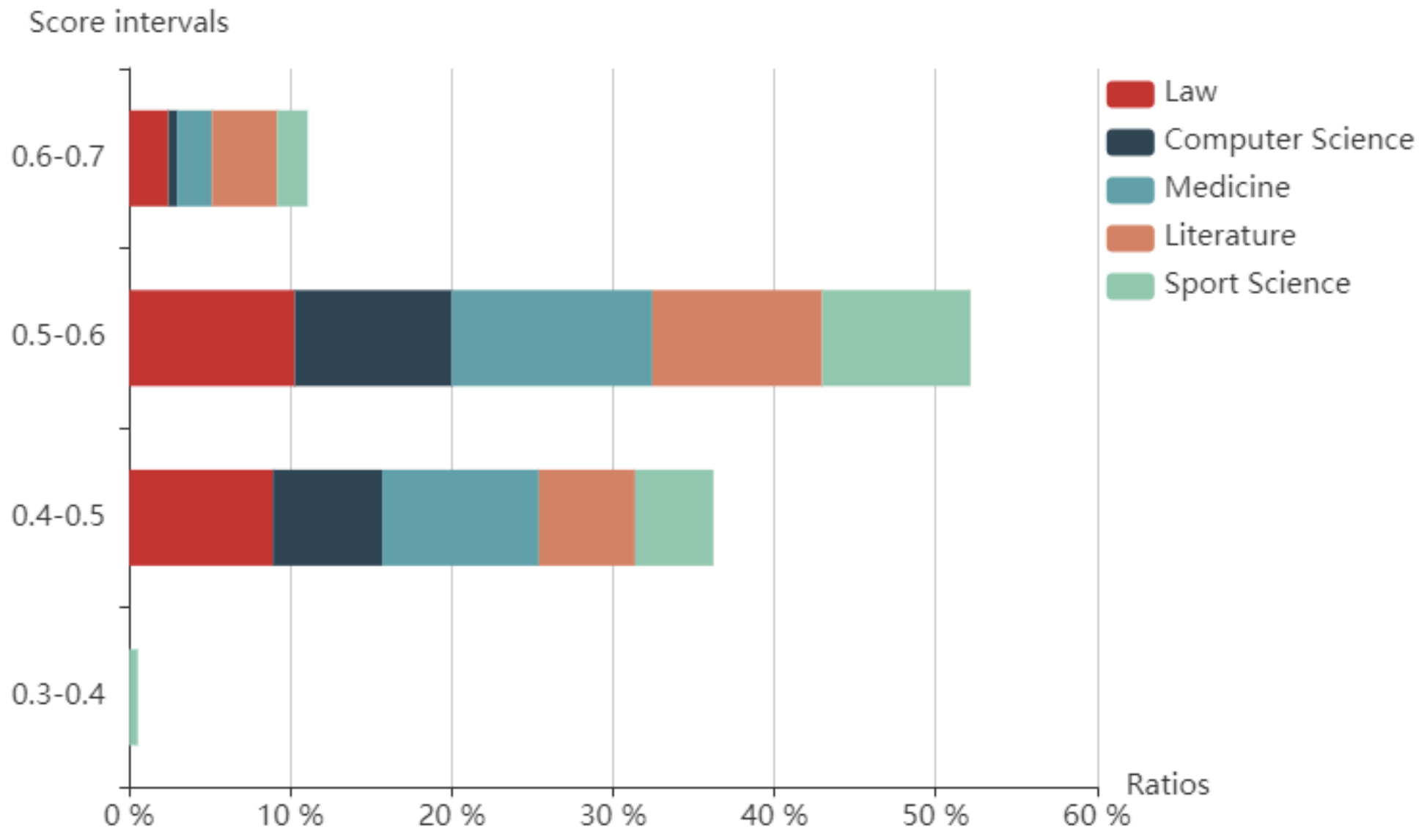

**Figure 12.** Distributions of book impact scores

# 6 Discussion

## *6.1 Comparative analysis with other evaluation methods*

This paper measured book impacts via integrating multiple evaluation resources including both internal and external evaluation resources of books. Compared with evaluation manually, book evaluation based on evaluation system can assess the impact of large numbers of books more quickly, reduce the cost of book evaluation research and shorten the evaluation cycle.

Compared with assessment research based on a single evaluation resource, this method can obtain the evaluation basis from more dimensions and more types of user groups, including book authors, researchers, ordinary readers and various institutional users (e.g. libraries). We conducted correlation analysis between expert evaluation scores and impact scores based on a single evaluation source, the correlation results are shown in Table 5. We can see from Table 5 that impact scores based on all four evaluation sources are significantly correlated with expert evaluation scores. It indicates that the four types of resources are reliable book impact evaluation resources, which can be used to measure different dimensions of book impact. However, the four correlation coefficients in Table 5 are lower than the correlation coefficients based on comprehensive evaluation (0.631 and 0.715). Hence, we can conclude that although the single evaluation source can be used to evaluate the impacts of books, the evaluation results are not comprehensive. The evaluation results obtained by integrating resources can overcome the one-sidedness of evaluation based on a single source, and avoid the situation that the book impact cannot be evaluated when lacking the certain dimension of evaluation data. More importantly, in some cases, users do not have a clear evaluation purpose or tendency. Thus, they are not sure which evaluation source is the most reliable basis for book selection, while comprehensive evaluation results can provide effective references for users, so as to effectively deal with such "evaluation cold start" phenomenon.

**Table 5.** Correlations between book impact scores based on single source and expert evaluation scores

| **Correlation** | | **Impact scores based on book content** | **Impact scores based on book review** | **Impact scores based on book citation** | **Impact scores based on book usage** |
|---|---|---|---|---|---|
| **Expert evaluation scores** | Computer science | 0.114* | 0.440** | 0.141* | 0.531** |
| | Literature | 0.103* | 0.531** | 0.159* | 0.269** |

Note: **. Significant at p=0.01, *. Significant at p=0.05

A noteworthy phenomenon is that for the four primary metrics, the metric weight of book content is slightly higher than the other three primary evaluation metrics, while the correlation coefficient between the impact scores based on book content and the expert evaluation scores is lower than other metrics. This may be related to the metrics delivered from the book content, that is, the TOC depth and TOC breadth. Existing studies have proved that the depth and breadth of books can be used to evaluate the impacts of books, but it is often difficult for book authors to balance the two (Zhang & Zhou, 2020). In other words, books with higher depth values are often difficult to get higher breadth values. We conducted correlation analysis between the TOC depth and TOC breadth, and the two metrics were highly negatively correlated (-0.820). Therefore, we can roughly convert the two metrics. Equation (20) shows the calculation of the comprehensive impact scores and conversion of the two secondary metrics extracted from book content.

$$\begin{aligned} Score_i &= S_{content_i} + S_{review_i} + S_{citation_i} + S_{usage_i} \\ &= w_{depth} * NorS_{TOCdepth_i} + w_{breadth} * NorS_{TOCbreadth_i} + S_{review_i} + S_{citation_i} + S_{usage_i} \\ &\cong w_{depth} * NorS_{TOCdepth_i} + w_{breadth} * \left(-k * NorS_{TOCdepth_i}\right) + S_{review_i} + S_{citation_i} + S_{usage_i} \\ &= \left(w_{depth} - k * w_{breadth}\right) * NorS_{TOCdepth_i} + S_{review_i} + S_{citation_i} + S_{usage_i} \end{aligned} \quad (20)$$

Where, $S_{content_i}$ is the impact scores based on book content of the book $i$, $S_{review_i}$, $S_{citation_i}$ and $S_{usage_i}$ are impact scores based on other three sources. $w_{depth}$ and $w_{breadth}$ are weights of the TOC depth and TOC breadth, $NorS_{TOCdepth_i}$ and $NorS_{TOCbreadth_i}$ denote normalized scores of the two metrics about book $i$, $k$ means the conversion coefficient of the two metrics. It can be seen from equation (20) that the high negative correlation between the two metrics weakens the weight of the primary metric (i.e. book content), and eventually leads to the weaker correlation between the impact scores based on book content and the comprehensive scores.

In addition, book impact evaluation based on the evaluation system can provide users with fine-grained analysis results, so as to support the decision-making of users from different groups. We take the book *Sweeping up fallen leaves for winter* as an example, the fine-grained analysis results are shown in Appendix C. From Appendix C we can see impact score of the book is ranked as 6 in this paper. In terms of book contents, the ranking of TOC depth is in the middle, while the ranking of TOC breadth is relatively low. We can conclude that the depth of the book is general and the scope of content is relatively small. In terms of book reviews, the book has many positive reviews and negative reviews, and 82% reviews are positive. Meanwhile, most users give 4-star or 5-star ratings for the book. It reveals that most users hold a positive attitude towards the book. In addition, the most satisfied and dissatisfied aspects are *printing* and *price*, while the most concerned and least concerned aspects are *content* and *font*. It indicates that satisfaction of *content* that users pay most attention to needs to be improved. For book citations, the ranking of citation frequency and citation literature depth is low, while citation literature breadth is high. It indicates that the book is less cited, while the topics of citations are diverse. Meanwhile, the book is most cited for use. In terms of book uses, this book has a large number of library holdings, and is collected by libraries in five countries around the world. The USA has the largest holding number of the book, followed by China. In conclusion, based on the analysis of multi-source evaluation data, we can get fine-grained evaluation results about books, and such results are difficult to obtain based on a single evaluation resource. In addition, the book impact evaluation results in structured rich text form in Appendix C can help users understand books more comprehensively and quickly, which is also the original intention of book impact evaluation research.

### *6.2 Book impact assessment based on users' diversified evaluation demands*

For users who have clear evaluation purposes (or evaluation needs), we can not only provide comprehensive evaluation results with detailed information, but also provide evaluation results based on specific evaluation resources according to users' different demands. This also reflects the advantages of the comprehensive evaluation system, that is, the differentiated combination of evaluation resources can adapt to the diversified and personalized evaluation tasks. For example, for users who want to refer to the previous purchase opinions or attitudes by existing users for book selection, we can provide them with book impact results based on book reviews, as shown in Table 6.

**Table 6.** Book impact assessment based on book reviews

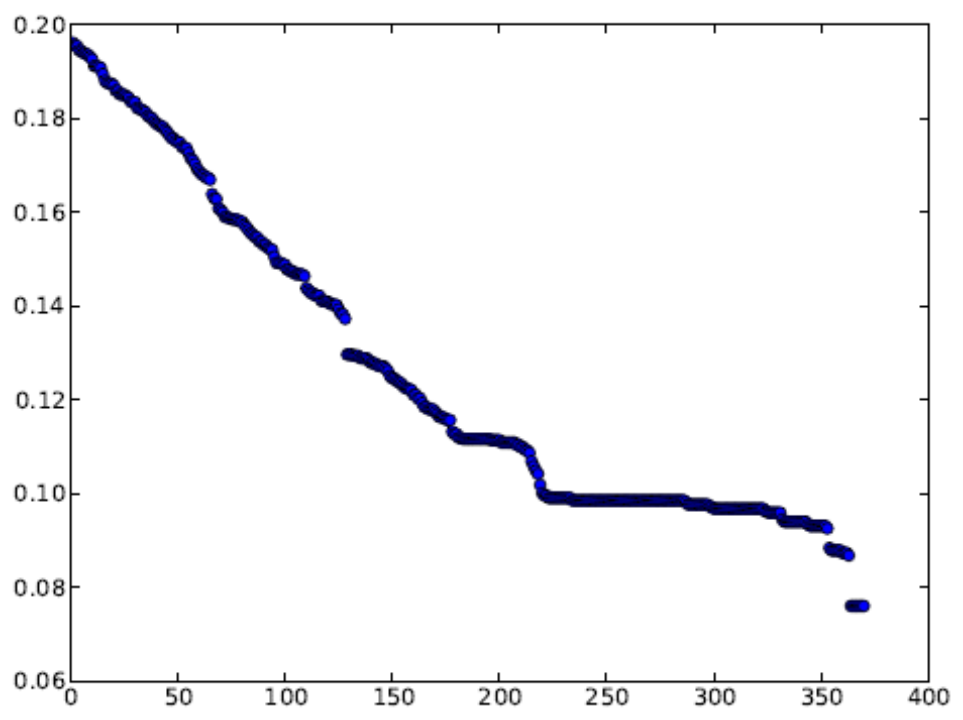

Book impact scores based on book reviews

| Rank | ISBN | Title | Discipline |
|---|---|---|---|
| 1 | 9787508633893 | *My Life* | Sport science |
| 2 | 9787108025371 | *Sweeping up fallen leaves for winter* | Law |
| 3 | 9787505732025 | *Memory is a light pain* | Literature |
| 4 | 9787532553129 | *Nalan's Poems* | Literature |
| 5 | 9787020102990 | *From the Seine to Firenze* | Literature |
| … | … | … | … |

For academic institutions, which pay more attention to the academic impacts of books, we can calculate impacts of books based on books' citation information, as shown in Table 7. Such book evaluation results can provide support for academic institutions to assist experts with awarding books, so as to improve the evaluation efficiency and reduce the award cost.

For libraries, they often need to consider the global library holdings and sales of books for book selections. Therefore, impact evaluation results based on book uses are often needed, as shown in Table 8. Based on such book impact assessment results, the libraries can quickly identify the books that need to be added, and adjust the position of books, so as to better ensure the circulation of books and ensure the libraries' customer flow.

For scholars, book content information is important for book recommendation. Hereby, impact evaluation is often measured based on book contents. The assessment results are shown in Table 9.

When selecting or recommending books, especially massive books with similar topics, scholars can choose books more quickly.

**Table 7.** Book impact assessment based on book citations

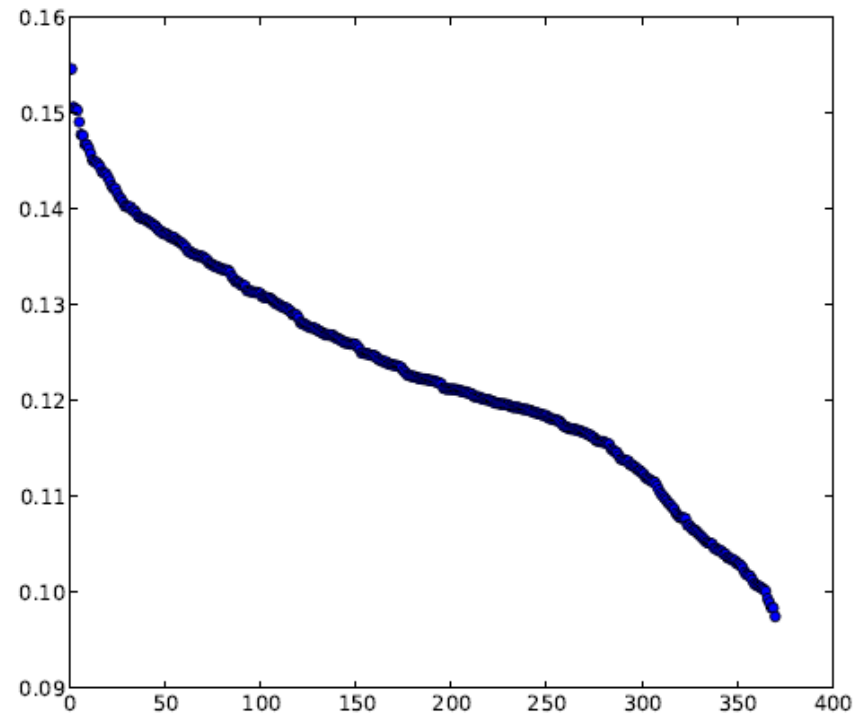

Book impact scores based on book citations

| Rank | ISBN | Title | Discipline |
|---|---|---|---|
| 1 | 9787807528876 | *On the vocabulary of Zhou Mi's notes* | Literature |
| 2 | 9787811210330 | *Zhongjing internal medicine* | Medicine |
| 3 | 9787811065497 | *Gynecopathy* | Medicine |
| 4 | 9787514606331 | *Four tragedies of Shakespeare* | Literature |
| 5 | 7308050467 | *Chinese and foreign literary selections of Yu Dafu* | Literature |
| … | … | … | … |

**Table 8.** Book impact assessment based on book usages

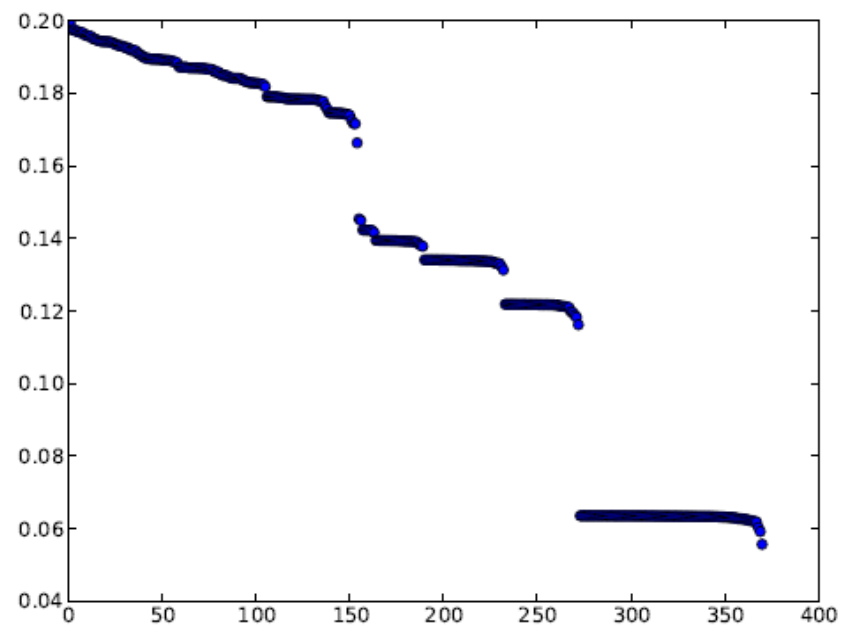

Book impact scores based on book usages

| Rank | ISBN | Title | Discipline |
|---|---|---|---|
| 1 | 9787306037602 | *Tips for healthy exercise you don't know* | Sport science |
| 2 | 7301094469 | *On Chinese traditional law: from the perspective of Chinese traditional studies* | Law |
| 3 | 7040220629 | *Selected lectures on drama of yuan, Ming and Qing Dynasties* | Literature |
| 4 | 9787117119726 | *Clinical parasitology laboratory* | Medicine |
| 5 | 9787301112496 | *Research on the theory of absolute property act and the legal system of real right* | Law |
| … | … | … | … |

**Table 9.** Book impact assessment based on book contents

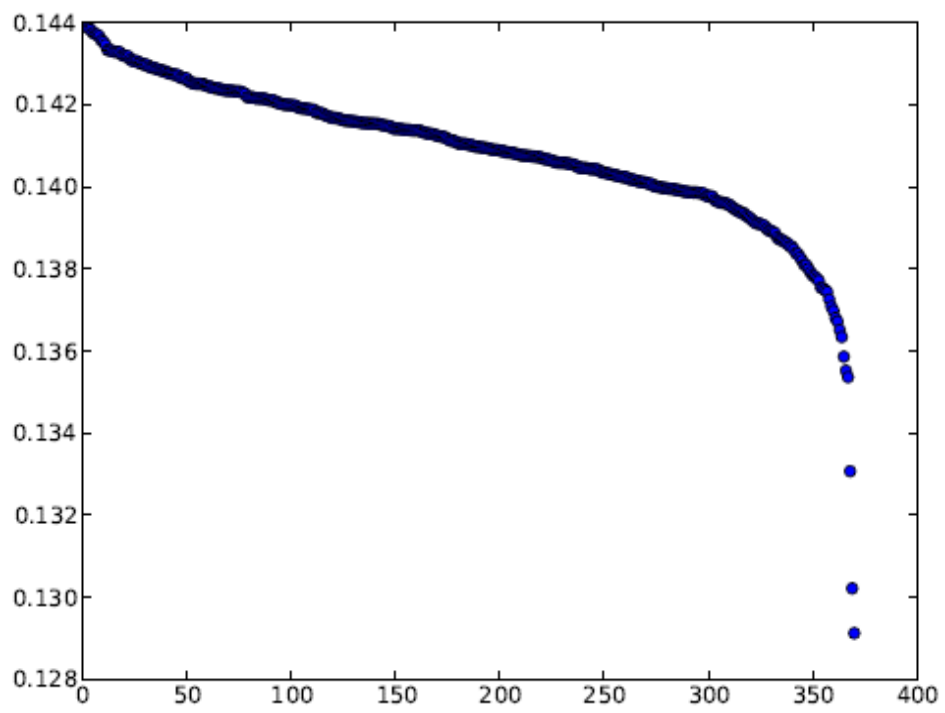

Book impact scores based on book contents

| Rank | ISBN | Title | Discipline |
|---|---|---|---|
| 1 | 9787514606331 | *Four tragedies of Shakespeare* | Literature |
| 2 | 9787301113028 | *Encyclopedia of law of Peking University: Economic Law* | Law |
| 3 | 7030128834 | *programmer's Manual of Visual foxpro8.0* | Computer science |
| 4 | 9787117134606 | *Handbook of rational use of antibiotics* | Medicine |
| 5 | 9787807087199 | *Appreciation Dictionary of 300 Yuan opera* | Literature |
| … | … | … | … |

In addition to providing evaluation results based on specific evaluation resources, users can also adjust the weight of each metric in the evaluation system according to their own needs, so as to obtain personalized evaluation results. However, it is worth noting that the adjustment of metric weights requires users to have a quite clear understanding of their evaluation needs.

Our study is subject to a few limitations. Firstly, due to the high cost of obtaining citation contents manually, data size in this paper is small. Hence, we will try to automatically detect the citation contents, so as to assess more books from more disciplines to further verify the reliability and feasibility of the evaluation system and methods proposed in this paper. Meanwhile, due to the sparsity of data (e.g. books' academic reviews published in journals), some evaluation resources are not included in the evaluation system of this paper. In the future, we need to explore the acquisition and analysis of such data, so as to improve the evaluation system. Secondly, in the process of integrating different resources, the quality difference of multiple evaluation resources also needs to be considered (Zhang et al., 2019). Measuring the data quality of different evaluation sources and screening reliable evaluation data is also a research direction of subsequent optimization. Meanwhile, it is necessary to integrate the evaluation data of the same evaluation resource in different platforms to avoid the evaluation error caused by a single platform. Lastly, this paper selected four evaluation resources from internal and external dimensions of books. However, there are still unidentified resources that can also be used to evaluate the impact of books. Therefore, in the follow-up study, we will excavate more reliable evaluation sources to improve the evaluation metric system.

## 7 Conclusion

This paper constructed an evaluation system for book impact and provided a comprehensive impact evaluation result. Meanwhile, users can integrate the required evaluation metrics according to different evaluation purposes and demands.

In answer to the first research question, the importance of metrics from the four resources is similar, while the weights of metrics extracted from book content are slightly higher. These evaluation metrics measure the impacts of books from different dimensions and play a complementary role in the impact evaluation process.

Regarding the second research question, the multi-source book impact assessment system does seem to be valuable for the book impact assessment. Meanwhile, assessment results based on the evaluation system can provide more detail information for different types of users and meet diverse users' evaluation needs.

Addressing the third research question, there are substantial differences between books published in different disciplines. In the book selection, recommendation and other related activities, it is necessary to fully consider the disciplinary differences of books.

In conclusion, book impacts measured based on the evaluation system can not only provide comprehensive evaluation results for users, but also obtain personalized evaluation results according to the evaluation needs of users. Meanwhile, this paper provides supplementary information for existing books evaluation, and it is suitable for various evaluation scenarios.

## Acknowledgement

This work is supported by the National Social Science Fund Project (No. 19CTQ031).

## Appendix A

# Questionnaire of assessment metrics about book impact

Dear scholars:

We are conducting research about book impact assessment. We have analyzed related works about book impact assessment, and a preliminary assessment system is structured (as shown in the following figure).

In order to improve the assessment system, please give your valuable opinion about importance of following assessment metrics. Assessment system includes four first-grade metrics: book reviews, book contents, book citations, book usages. Each first-grade metric has corresponding second-grade metrics.  Please assess the importance of metrics at all grades.

**1: Very unimportant  2: Not important  3: General importance  4: Relative important  5: Very important**

Thank you for your support and cooperation.

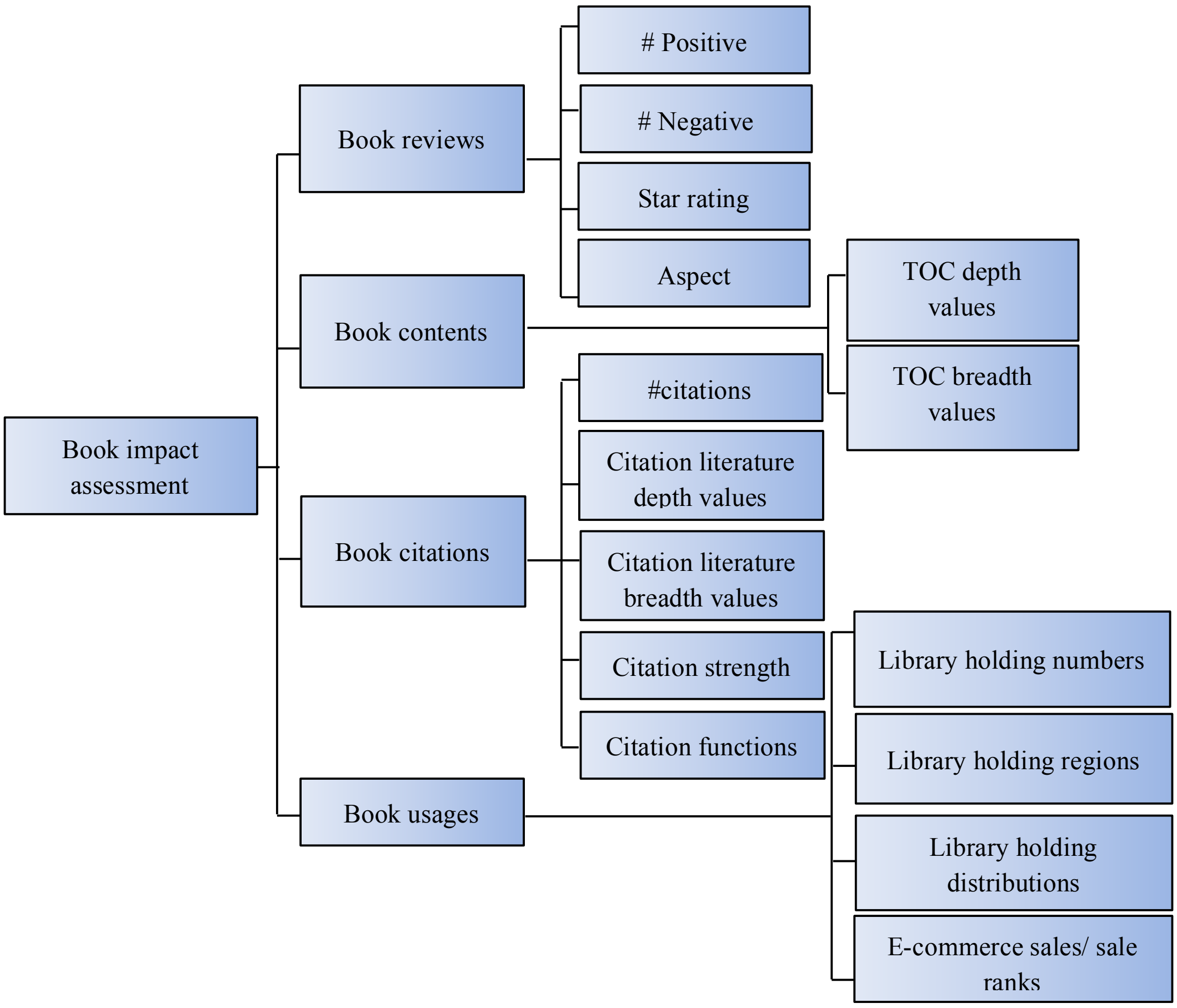

## Part1: Your basic information

**Major:** ____________________  **E-mail:** ____________________

**Your educational background:**

- ○ Below the undergraduate level
- ○ Undergraduate
- ○ Master
- ○ Doctorate and above

**Your educational background:**

- ○ Assistant professor
- ○ Associate Professor
- ○ Professor
- ○ Other

## Part2: Importance of assessment metrics

### Q2: The importance of first-grade indexes:

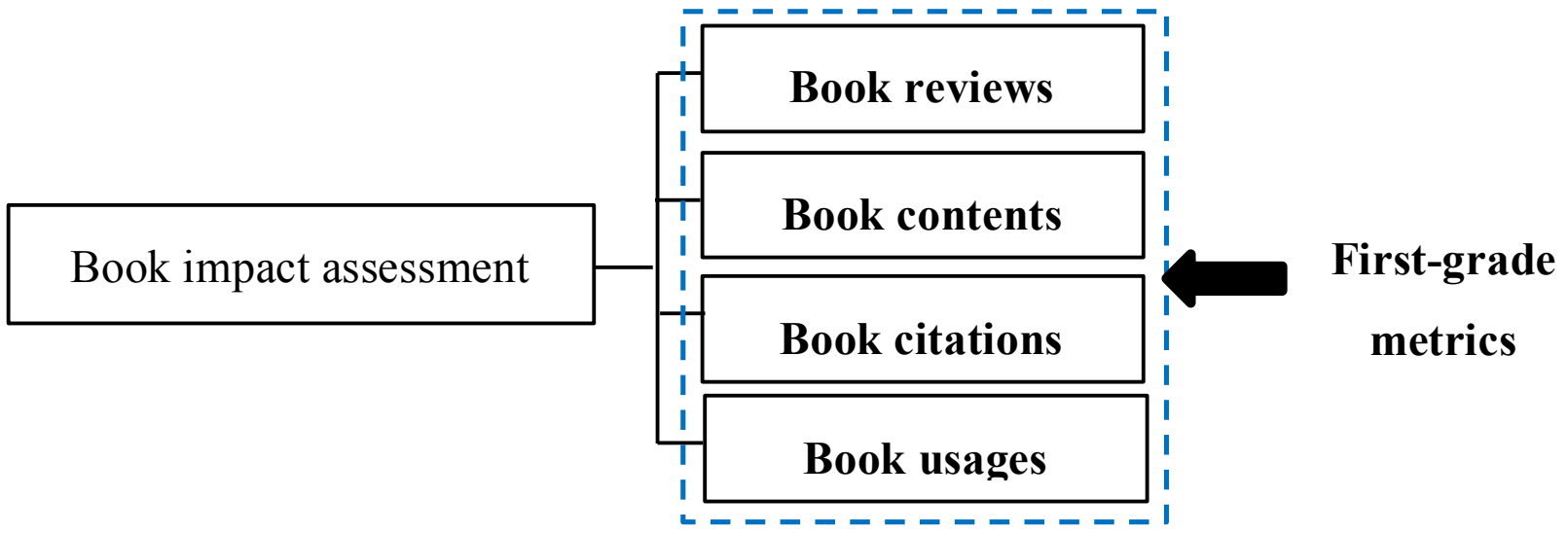

| | Very unimportant | | → | | Very important |
|---|---|---|---|---|---|
| Book reviews: | □ 1 | □ 2 | □ 3 | □ 4 | □ 5 |
| Book contents: | □ 1 | □ 2 | □ 3 | □ 4 | □ 5 |
| Book citations: | □ 1 | □ 2 | □ 3 | □ 4 | □ 5 |
| Book usages | □ 1 | □ 2 | □ 3 | □ 4 | □ 5 |

### Q3: The importance of second-grade indexes about book reviews:

**# Positive reviews**: Number of positive reviews about this book given by users
**# Negative reviews**: Number of negative reviews about this book given by users
**Star rating**: Star ratings given by users
**Aspect satisfactions**: Users' satisfaction about book aspects (aspects refer to *price*, *printing* etc.)

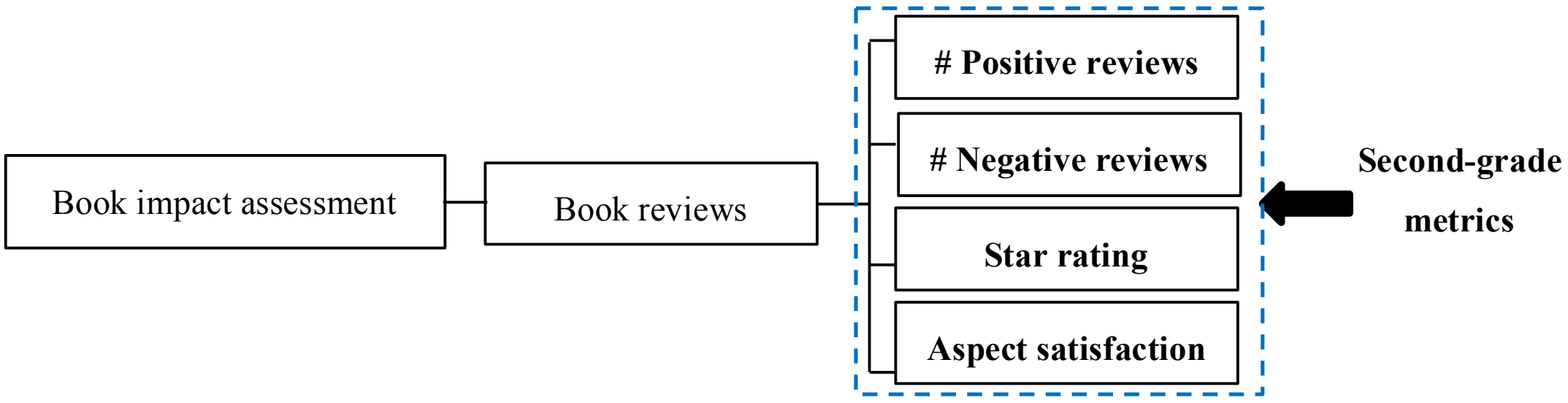

| | Very unimportant | | → | | Very important |
|---|---|---|---|---|---|
| # positive reviews: | □ 1 | □ 2 | □ 3 | □ 4 | □ 5 |
| # negative reviews: | □ 1 | □ 2 | □ 3 | □ 4 | □ 5 |
| Star rating: | □ 1 | □ 2 | □ 3 | □ 4 | □ 5 |
| Aspect satisfactions: | □ 1 | □ 2 | □ 3 | □ 4 | □ 5 |

### Q4: The importance of second-grade indexes about book contents:

**TOC depth values:** Depth of books reflected by books' tables of contents. Higher depth value of books means books introduced deeper theory, technology, etc.

**TOC breadth values**: Breadth of books reflected by books' tables of contents. Higher breadth value of books means book involved a wider range of knowledge, and introduced more theory, technology, etc.

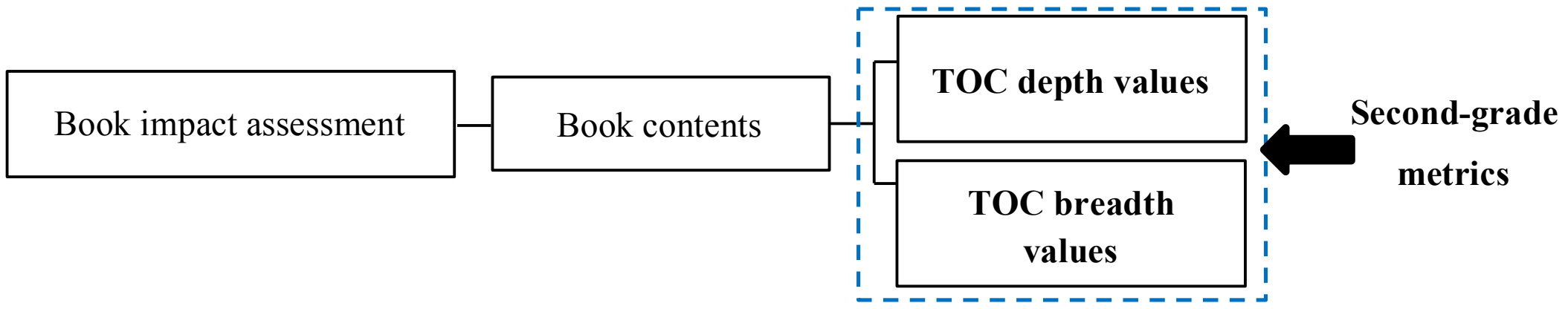

| | Very unimportant | | → | | Very important |
|---|---|---|---|---|---|
| TOC depth values: | □ 1 | □ 2 | □ 3 | □ 4 | □ 5 |
| TOC breadth values: | □ 1 | □ 2 | □ 3 | □ 4 | □ 5 |

**Q5: The importance of second-grade indexes about book citations:**

**#citations**: Citation frequency of this book
**Citation literature depth values**: Depth of the book reflected by literatures which cited this book
**Citation literature breadth values**: Breadth of the book reflected by literatures which cited this book
**Citation strength**: Citation times of this book in one literature by analyzing citation context
**Citation functions**: Citation function refers to the use of this book cited by other literatures, e.g. background citation, method citation etc.

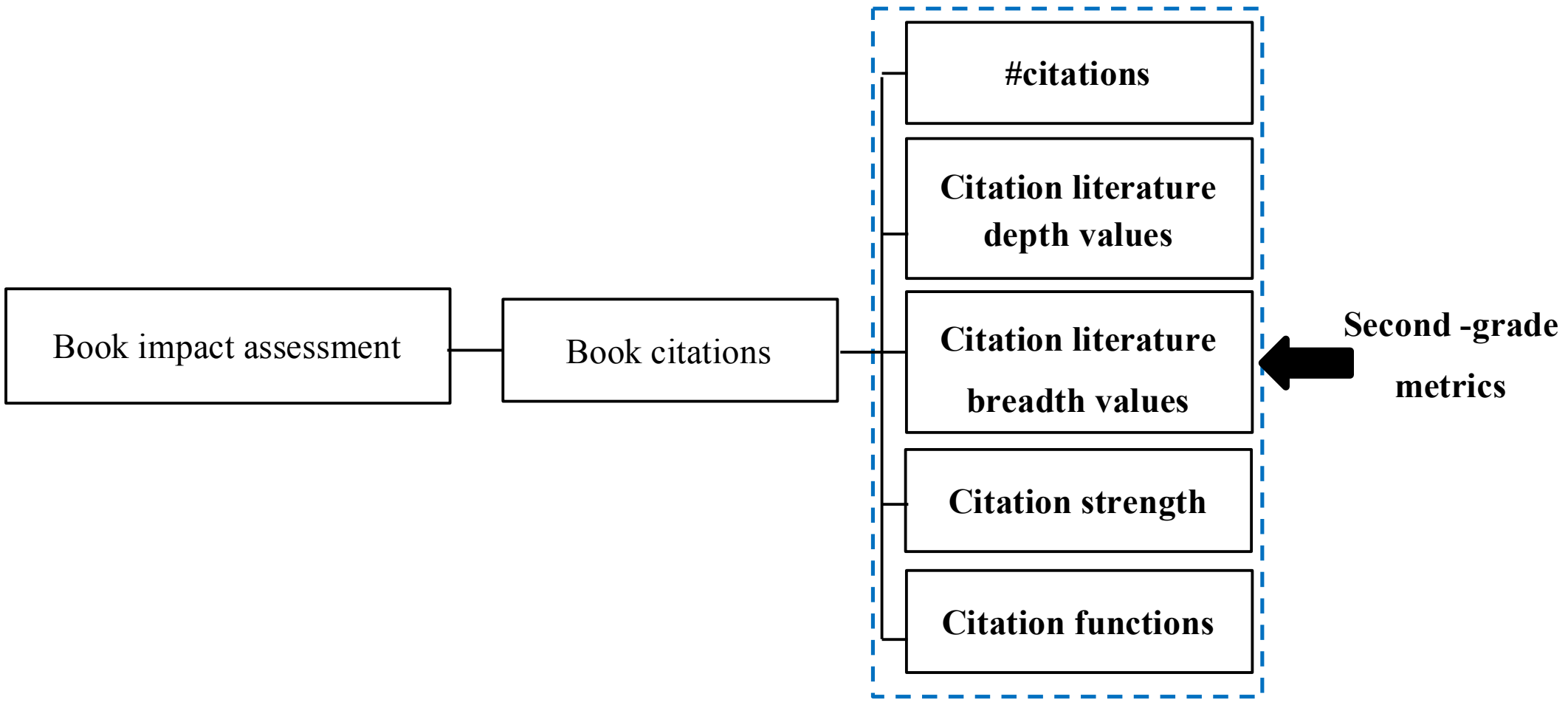

| | Very unimportant | | → | | Very important |
|---|---|---|---|---|---|
| #citations: | □ 1 | □ 2 | □ 3 | □ 4 | □ 5 |
| Depth values: | □ 1 | □ 2 | □ 3 | □ 4 | □ 5 |
| Breadth values: | □ 1 | □ 2 | □ 3 | □ 4 | □ 5 |
| Citation strength: | □ 1 | □ 2 | □ 3 | □ 4 | □ 5 |
| Citation functions: | □ 1 | □ 2 | □ 3 | □ 4 | □ 5 |

**Q6: The importance of second-grade indexes about book usages:**

**Library holding numbers**: Total number of collections about this book in various libraries around the world

**Library holding regions**: Total number of library regions that collect this book

**Library holding distributions:** Holding distributions of this book in various libraries around the world

**E-commerce sales/ sale ranks**: The sales of books on e-commerce website

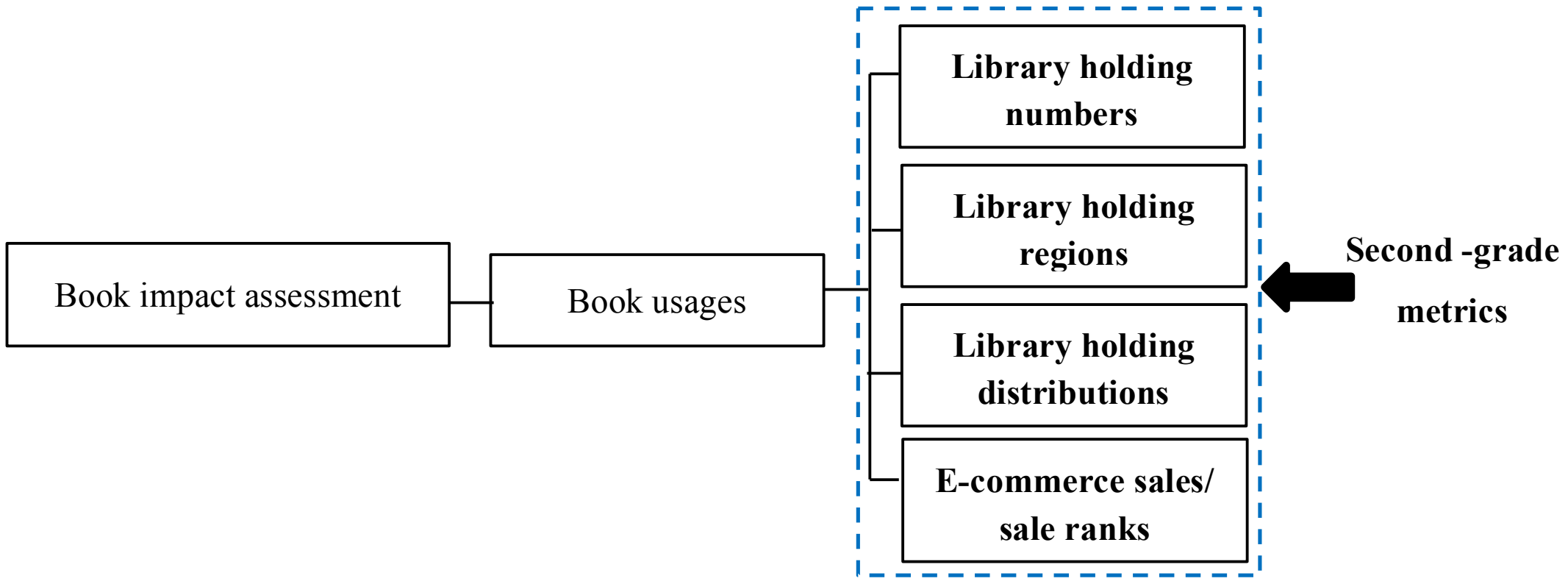

| | Very unimportant | | | | Very important |
|---|---|---|---|---|---|
| Library holding numbers: | □ 1 | □ 2 | □ 3 | □ 4 | □ 5 |
| Library holding regions: | □ 1 | □ 2 | □ 3 | □ 4 | □ 5 |
| Library holding distributions: | □ 1 | □ 2 | □ 3 | □ 4 | □ 5 |
| E-commerce sales/ sale ranks: | □ 1 | □ 2 | □ 3 | □ 4 | □ 5 |

## Appendix B

# Questionnaire of the impacts of books in literature

Dear scholars:

We are conducting research about book impact assessment. You are invited to assess the impacts of books in the following five domains of literature. You can make a comprehensive assessment according to books' citations, reviews, sales, library holdings etc., and then give the impact score grades of books.

**1: Low impact 2: Relative low impact 3: General impact 4: Relative high impact 5: High impact**

Thank you for your support and cooperation.

---

## Part1: Your basic information

**Major:** ___________________ **E-mail:** ___________________

**Your educational background:**

- ○ Below the undergraduate level
- ○ Undergraduate
- ○ Master
- ○ Doctorate and above

## Part2: Book impact assessment

*Q2: Books in the domain of literature research*

| ID | Title | Authors | Publishers |
|---|---|---|---|
| 1 | History of Chinese literature | Lin Geng | Tsinghua University Press, 2009 |
| 2 | A brief history of world literature | Li Mingbin | Peking University Press, 2002 |
| 3 | Japanese elegance | Onishi Yoshinori | Beijing Jiban books Co., Ltd., 2012 |
| 4 | Psychology of contemporary literature and art | Jinyuanpu | China Renmin University Press, 2009 |
| 5 | History of fiction: Theory and Practice | Chen Pingyuan | Peking University Press, 2010 |
| 6 | The foundation of modern literature | Zhang Fugui, Wang Xueqian, Liu Zhongshu | Peking University Press, 2009 |
| 7 | History of ancient Chinese Literature | Guo Yuheng | Shanghai Classics Publishing House, 1998 |

*(click on the title of the book below to get more information about the book)*

| Title | Low impact | | → | | High impact |
|---|---|---|---|---|---|
| History of Chinese literature | □1 | □ 2 | □3 | □4 | □5 |
| A brief history of world literature | □1 | □ 2 | □3 | □4 | □5 |
| Japanese elegance | □1 | □ 2 | □3 | □4 | □5 |
| Psychology of contemporary literature and art | □1 | □ 2 | □3 | □4 | □5 |
| History of fiction: Theory and Practice | □1 | □ 2 | □3 | □4 | □5 |
| The foundation of modern literature | □1 | □ 2 | □3 | □4 | □5 |
| History of ancient Chinese Literature | □1 | □ 2 | □3 | □4 | □5 |

*Q3: Books in the domain of novel*

| ID | Title | Authors | Publishers |
|---|---|---|---|
| 1 | The true story of Ah Q | Lu Xun | China Overseas Chinese press, 2013 |
| 2 | A woman with flowers in her arms | Mo Yan | Shanghai Literature and Art Publishing House, 2012 |
| 3 | Comments on a dream of Red Mansions | Wang Guowei, Cai Yuanpei | Shanghai Classics Publishing House, 2011 |
| 4 | The Peach Blossom Fan annotated by Liang Qichao | Kong Shangren, Liang Qichao | Phoenix publishing house, 2011 |
| 5 | On the version of dream of Red Mansions | Lin Guanfu | Culture and Art Press, 2007 |
| 6 | Red Mansions in the wind | Miao huaiming | Zhonghua Book Company, 2006 |

*(click on the title of the book below to get more information about the book)*

| Title | Low impact | | → | High impact | |
|---|---|---|---|---|---|
| The true story of Ah Q | □ 1 | □ 2 | □ 3 | □ 4 | □ 5 |
| A woman with flowers in her arms | □ 1 | □ 2 | □ 3 | □ 4 | □ 5 |
| Comments on a dream of Red Mansions | □ 1 | □ 2 | □ 3 | □ 4 | □ 5 |
| The Peach Blossom Fan annotated by Liang Qichao | □ 1 | □ 2 | □ 3 | □ 4 | □ 5 |
| On the version of dream of Red Mansions | □ 1 | □ 2 | □ 3 | □ 4 | □ 5 |
| Red Mansions in the wind | □ 1 | □ 2 | □ 3 | □ 4 | □ 5 |

*Q4: Books in the domain of poetry and drama*

| ID | Title | Authors | Publishers |
|---|---|---|---|
| 1 | Nalan's poetry and lyrics | Zhang caozhen, Nalanxingde | Shanghai Literature and Art Publishing House, 2009 |
| 2 | Four tragedies of Shakespeare | William Shakespeare | China Pictorial press, 2013 |
| 3 | Recite progress | Zhang Benyi | Guangxi Normal University Press, 2013 |
| 4 | On the original poem | Ye Xie, Shen Deqian | Phoenix publishing house, 2010 |
| 5 | A study on the vocabulary of Zhoumi notes | Yang Guan | Bashu publishing house, 2011 |
| 6 | Lectures on famous Ci Poems of Tang and Song Dynasties | Wang Zhaopeng | Guangxi Normal University, 2006 |
| 7 | Xi Murong's classic works | Xi Murong | Contemporary world press, 2007 |
| 8 | Collection of Ming Dynasty folk songs | Zhou Yubo, Chen Shulu | Nanjing Normal University Press, 2009 |
| 9 | Hamlet's problem | Zhang Pei | Peking University Press, 2006 |

*(click on the title of the book below to get more information about the book)*

| Title | Low impact | | → | | High impact |
|---|---|---|---|---|---|
| Nalan's poetry and lyrics | □ 1 | □2 | □3 | □4 | □5 |
| Four tragedies of Shakespeare | □ 1 | □2 | □3 | □4 | □5 |
| Recite progress | □ 1 | □2 | □3 | □4 | □5 |
| On the original poem | □ 1 | □2 | □3 | □4 | □5 |
| A study on the vocabulary of Zhoumi notes | □ 1 | □2 | □3 | □4 | □5 |
| Lectures on famous Ci Poems of Tang and Song Dynasties | □ 1 | □2 | □3 | □4 | □5 |
| Xi Murong's classic works | □ 1 | □2 | □3 | □4 | □5 |
| Collection of Ming Dynasty folk songs | □ 1 | □2 | □3 | □4 | □5 |
| Hamlet's problems | □ 1 | □2 | □3 | □4 | □5 |

*Q5: Books in the domain of prose*

| ID | Title | Authors | Publishers |
|---|---|---|---|
| 1 | Memory is a light pain | Long Yingtai, Jiang Xun | China Friendship Publishing Company, 2013 |
| 2 | May you embrace the world warmly | Bi Shumin | Jiangsu literature and Art Publishing House, 2013 |
| 3 | Li Ao's love letters | Li Ao | Time literature and Art Press, 2012 |
| 4 | Sleep empty | Annie baby (Qingshan) | Beijing October literature and Art Publishing House, 2013 |
| 5 | Along the Seine to Firenze | Huang Yongyu | People's Literature Press, 2014 |

*(click on the title of the book below to get more information about the book)*

| Title | Low impact | | → | | High impact |
|---|---|---|---|---|---|
| Memory is a light pain | □ 1 | □ 2 | □ 3 | □ 4 | □ 5 |
| May you embrace the world warmly | □ 1 | □ 2 | □ 3 | □ 4 | □ 5 |
| Li Ao's love letters | □ 1 | □ 2 | □ 3 | □ 4 | □ 5 |
| Sleep empty | □ 1 | □ 2 | □ 3 | □ 4 | □ 5 |
| Along the Seine to Firenze | □ 1 | □ 2 | □ 3 | □ 4 | □ 5 |

*Q6: Books in the domain of history*

| ID | Title | Authors | Publishers |
|---|---|---|---|
| 1 | The Rommel Papers | Liddle Hart | Democracy and construction press, 2015 |
| 2 | Military diary | Xie Bingying | Jiangsu literature and Art Publishing House, 2010 |
| 3 | Yu Qiuli and the oil war | Chen Daokuo | PLA literature and Art Publishing House, 2009 |

*(click on the title of the book below to get more information about the book)*

| Title | Low impact | | → | | High impact |
|---|---|---|---|---|---|
| The Rommel Papers | □ 1 | □ 2 | □ 3 | □ 4 | □ 5 |
| Military diary | □ 1 | □ 2 | □ 3 | □ 4 | □ 5 |
| Yu Qiuli and the oil war | □ 1 | □ 2 | □ 3 | □ 4 | □ 5 |

## Appendix C

**Fine-grained analysis of impact scores of *Sweeping up fallen leaves for winter***

| ISBN | Title | Disciplines | Impact rank |
|---|---|---|---|
| 9787108025371 | Sweeping up fallen leaves for winter | Law | 6 |

**Book contents**

| | |
|---|---|
| TOC depth rank: 191 | TOC breadth rank: 281 |

**Book citations**

| | |
|---|---|
| #citations rank: 356 | Citation literature depth rank: 370 |
| Citation literature breadth rank: 59 | |
| Citation strength rank: 142 | Citation function rank: 34 |

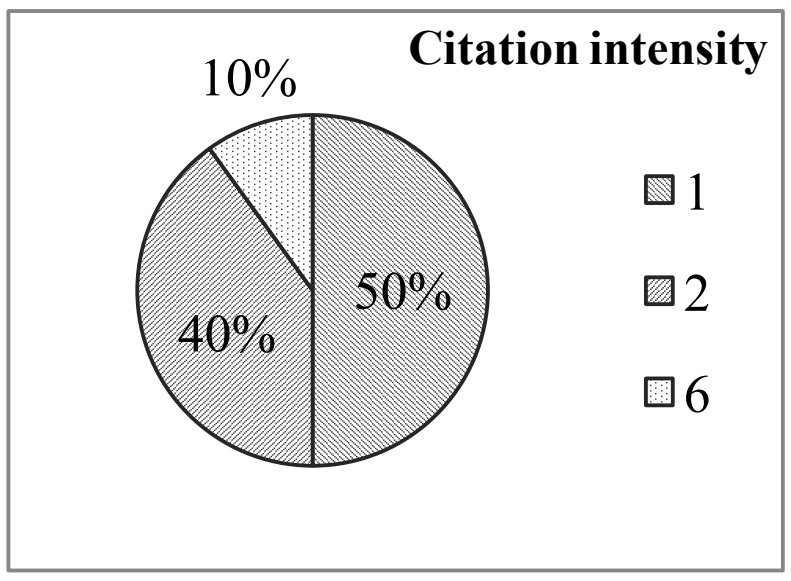

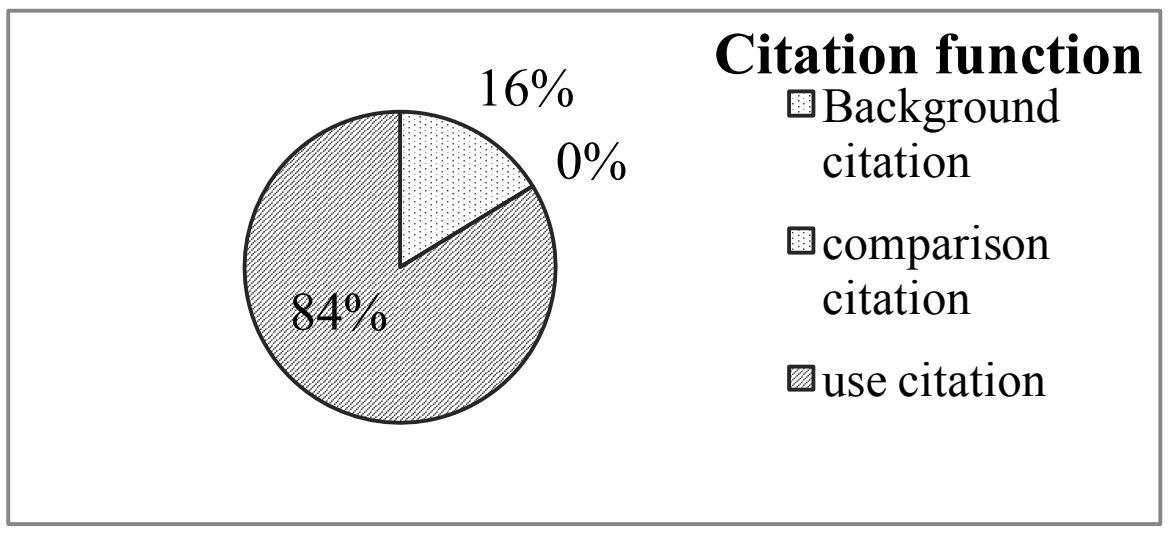

**Book usages**

| | |
|---|---|
| Holding number rank: 9 | Holding region rank: 10 |
| Holding distribution rank: 137 | E-commerce sale rank: 17 |

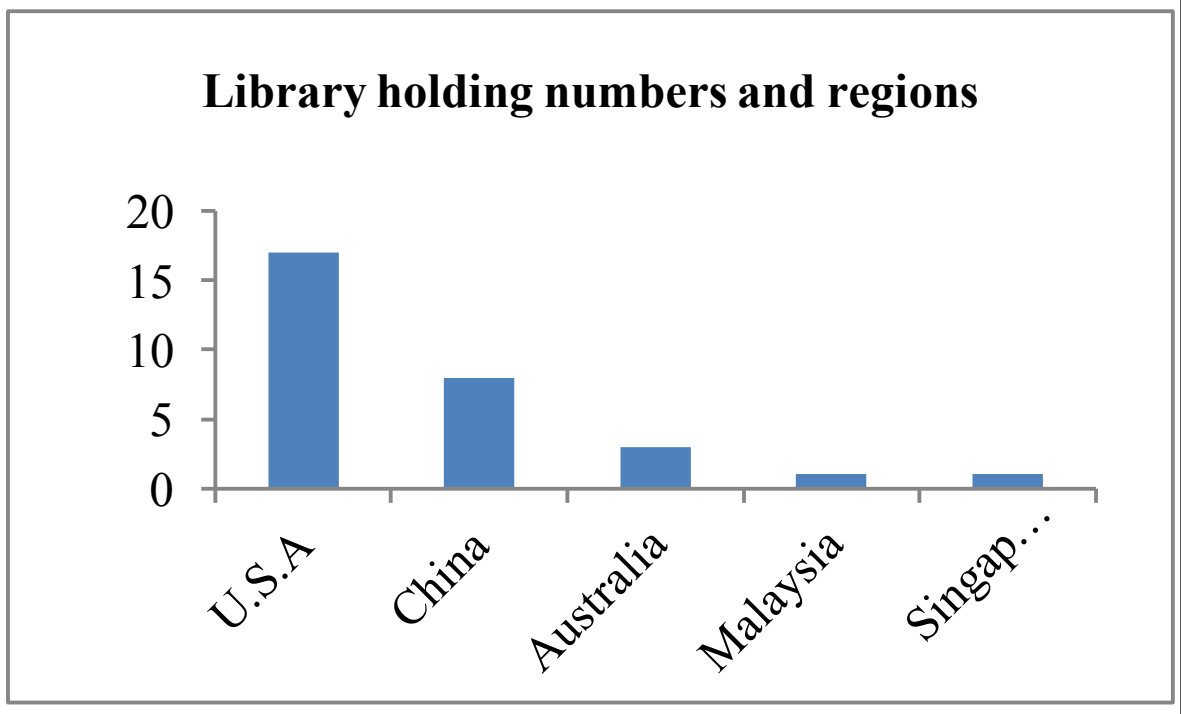

**Book reviews**

| | | |
|---|---|---|
| #positive reviews V.S. #negative reviews | #positive review rank: 6 | #negative review rank: 8 |

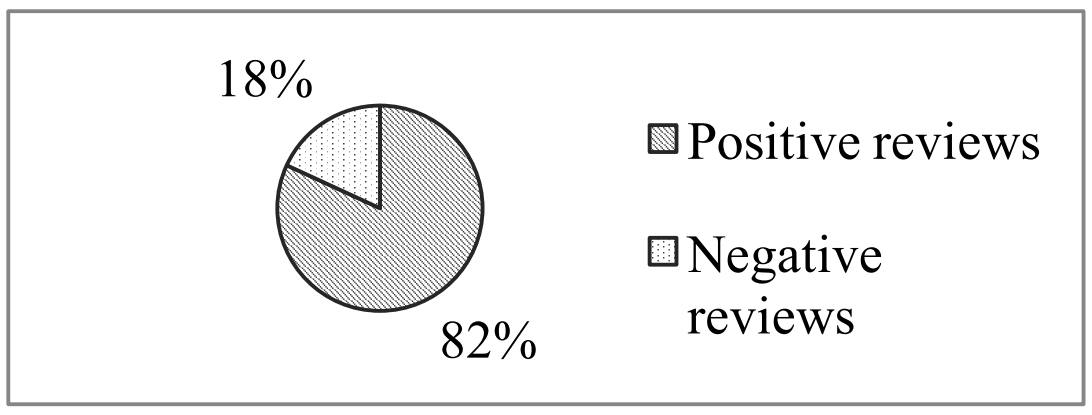

Star ratings

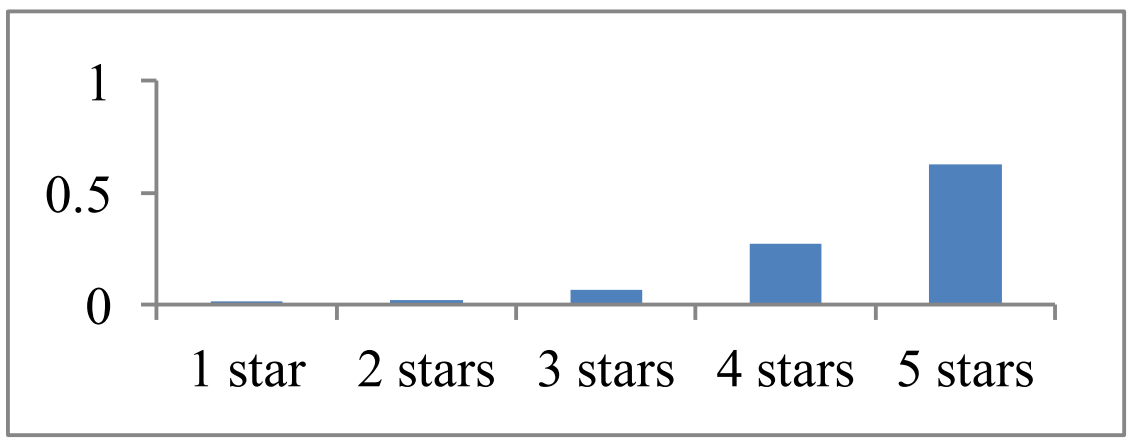

Aspect

| Most satisfied aspect: | Least satisfied aspect: |
|---|---|
| *Printing* | *Price* |

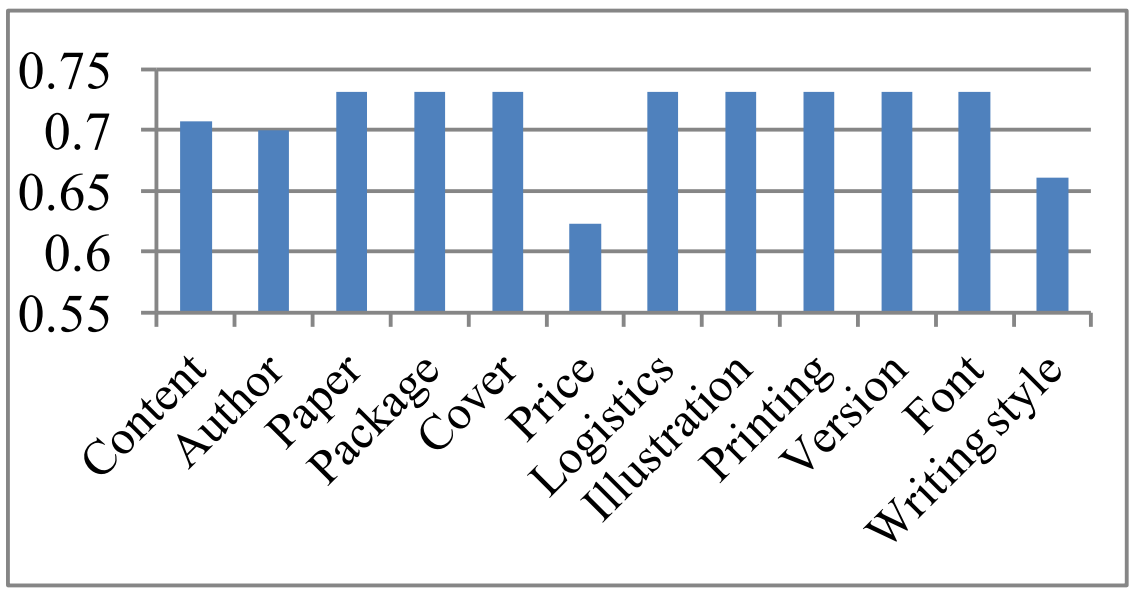

| Most concerned aspect: | Least concerned aspect: |
|---|---|
| *Content* | *Font* |

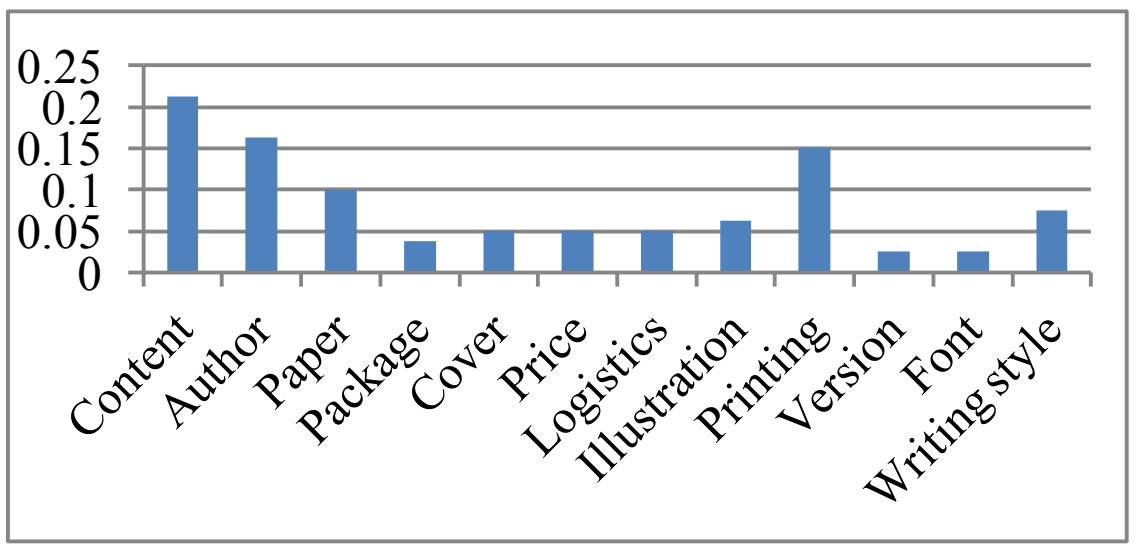